\documentclass[preprint,12pt]{elsarticle}

\usepackage[dvipsnames,svgnames,x11names, table,xcdraw]{xcolor}
\usepackage[version=3]{mhchem} 
\usepackage{caption}
\usepackage{lineno}
\usepackage{subcaption}
\usepackage{hyperref}
\usepackage{amssymb}
\usepackage{multirow}
\usepackage{amsmath}

\journal{Chemical Engineering Journal}

\begin{document}

\begin{frontmatter}



\title{Pooling solvent mixtures for solvation free energy predictions}


\affiliation[KUL]{organization={Department of Chemical Engineering, KU Leuven},
            addressline={Celestijnenlaan 200F},
            city={Leuven},
            postcode={3000},
            country={Belgium}}

\affiliation[MIT]{organization={Department of Chemical Engineering, Massuchusetts Institute of Technology},
            city={Cambridge},
            postcode={02139},
            state={MA},
            country={USA}}

\author[KUL]{Roel J. Leenhouts\corref{cor}}
\author[MIT]{Nathan Morgan\corref{cor}} 
\author[MIT]{Emad Al Ibrahim}
\author[MIT]{William H. Green}
\author[KUL]{Florence H. Vermeire\corref{cor1}}

\cortext[cor]{Contributed equally}
\cortext[cor1]{Corresponding author \newline \textit{Email address:} \url{florence.vermeire@kuleuven.be}}


\begin{abstract}

Solvation free energy is an important design parameter in reaction kinetics and separation processes, making it a critical property to predict during process development. In previous research, directed message passing neural networks (D-MPNN) have successfully been used to predict solvation free energies and enthalpies in organic solvents. However, solvent mixtures provide greater flexibility for optimizing solvent interactions than monosolvents. This work aims to extend our previous models to mixtures.
To handle mixtures in a permutation invariant manner we propose a pooling function; MolPool. With this pooling function, the machine learning models can learn and predict properties for an arbitrary number of molecules.
The novel SolProp-mix software that applies MolPool to D-MPNN was compared to state-of-the-art architectures for predicting mixture properties and validated with our new database of COSMOtherm calculations; BinarySolv-QM.
To improve predictions towards experimental accuracy, the network was then fine-tuned on experimental data in monosolvents.
To demonstrate the benefit of this transfer learning methodology, experimental datasets of solvation free energies in binary (BinarySolv-Exp) and ternary (TernarySolv-Exp) solvent mixtures were compiled from data on vapor-liquid equilibria and activity coefficients.
The neural network performed better than COSMOtherm calculations with an MAE of $0.25 \ kcal/mol$ and an RMSE of $0.37 \ kcal/mol$ for non-aqueous mixed solvents.
Additionally, the ability to capture trends for a varying mixture composition was validated successfully. Our model's ability to accurately predict mixture properties from the combination of \textit{in silico} data and pure component experimental data is promising given the scarcity of experimental data for mixtures in many fields.

\end{abstract}






\end{frontmatter}



\section{Introduction}

Research on thermochemical property prediction has recently surged because of the application of machine learning (ML) methods such as graph neural networks (GNNs) and transformers \cite{Chemprop, Winter, GNN_review, Gilmer, Barzilay}. These methods have been widely applied to thermochemical properties of both pure compounds and infinite-diluted mixtures, outperforming well established thermodynamic models. Accordingly, applications to numerous properties were shown, for example, activity coefficients \cite{Qin, RittigGibbsDuhem, gibbs_excess, Medina, hanna}, solubilities \cite{VermeireSolu}, viscosities \cite{BiloViscosities}, liquid densities \cite{FuelProperties}, and vapour pressures \cite{Felton}. These applications outperform renowned models such as UNIFAC \cite{UNIFAC} and COSMO-RS \cite{cosmors}. Recent research has studied infinite-diluted \cite{Medina, VermeireTL} and composition dependent mixtures \cite{Qin, RittigGibbsDuhem}, as well as temperature dependence \cite{MedinaTdep}.

As part of thermochemical property research, solvation free energy is actively investigated because solvation is of critical influence to the optimization of reactions in solution and the design of new separation processes \cite{VermeireSolu, VermeireTL, Boobier, Chung}. 
Reactions are influenced by the solvation free energy because it affects the activation energy of the reaction in solution through the Eyring equation, which in turn influences reaction rates and pathways.
Moreover, solvation properties, e.g. activity coefficients and solvation free energy, correlate to the solubility of gases, liquids, and solids. 
Solvation free energy predictors have focused on infinite-diluted binary systems of active pharmaceutical ingredients \cite{VermeireTL}; however, in industry, it is common practice to formulate solvent mixtures to increase the reaction rate and solubility compared to individual solvents \cite{qiu_synergistic_2019}. Solvent mixtures increase the exploration space of economical viable solvents because of the broader range of parameters that can be optimized to reach similar objectives as in the design of \textit{de novo} solvents. \textit{De novo} solvents are usually designed for specifically applications but often fail to take into account toxicological, economical, or synthesizable viability. The potential to formulate a solvent mixture is argued to be the single most useful idea of the famous Hansen solubility parameters \cite{Hansen_2012}. Therefore, an extension of machine learning methods for solvation free energy prediction to solvent mixtures is timely.

A variety of ML architectures for predicting properties of mixtures have been proposed \cite{Hanaoka}. Although there have been studies to predict activity coefficients that included compositional dependence \cite{Qin, RittigGibbsDuhem, Medina}, none of these proposed a method that can process an arbitrary number of components in the mixture in a permutation invariant way. In other words, they cannot reuse the same neural network for both binary and ternary mixtures and hence fail to include the information of these highly similar systems in a single neural network. Therefore, we adapt an architecture from Hanaoka and Kyohei \cite{Hanaoka} that is permutation invariant with respect to the number and order of components. This architecture uses a weighted averaging of individual component embeddings to pool these embeddings into an embedding for the mixture. This is analogous to the classical pooling function in message passing neural networks (MPNNs) that pools the atom embeddings into a molecular embedding \cite{PhysicalPooling}. This pooling function creates the same embedding when the input order of molecules is shuffled, which is useful as also the components in homogeneous mixtures are not naturally ordered.
Such pooling function also allows the model to train on information from single molecule property data and transfer this information to mixture predictions. This is important because single molecule data are often available in higher quantities and more molecular diversity. 

The goal of this work is the accurate and fast prediction of infinite dilute solvation free energies and enthalpies in mixed solvents for organic, neutral solutes. For this purpose, a quantum mechanics (QM) dataset was created considering solvation free energies and enthalpies in binary solvent mixtures, called BinarySolv-QM. Subsequently, a new network architecture was proposed and compared to state-of-the-art architectures. While several datasets are available for experimental solvation free energies in monosolvents, there is not an analogue for mixed solvents. To test the performance of this architecture on experimental solvation free energies in mixed solvents, new datasets were curated for binary solvents, BinarySolv-Exp, and ternary solvents, TernarySolv-Exp. Finally, the prediction of trends in varying solvent mole fractions were verified and temperature dependence for moderate temperatures was considered.

\section{Methods} 
\label{section:ddb}      

\subsection{The experimental and quantum chemical databases}
In this work, synthetic and experimental databases containing solvation free energies in solvent mixtures were compiled. Additionally, synthetic and experimental datasets for solvation free energies in monosolvents were used. For the details on these monosolvent datasets, we refer to previous work \cite{VermeireSolu}. With our synthetic datasets, we aim to cover a broad range of solutes to extend the applicability range of our model. The curation of experimental data was aimed at obtaining a dataset for real-world validation and comparison of the models.

\subsubsection{A quantum chemical database for binary solvent mixtures}

The synthetic dataset (BinarySolv-QM) of infinite dilution solvation free energies and enthalpies at $298 K$ in binary solvent mixtures was calculated based on the COSMO-RS theory \cite{cosmors}. Data generation was performed using the commercial software COSMOtherm \cite{Dassault}. For the output of the calculations, the "molar" framework was selected, which is defined as the calculation of the free energy that a solute molecule $i$ requires to go from an ideal gas state at one molar concentration to an ideal solution at one molar solute concentration.
The COSMO-surfaces used in this work were computed at the BP-TZVPD-FINE level of theory, \textit{i.e.} using a geometry which was optimized at the density functional theory BP-TZVP level, a single point energy calculation at the BP-def2-TZVPD level and a FINE cavity for the construction of the surface segments. Different conformations of the solute and solvents were included, generated according to the default algorithm implemented in COSMOconf \cite{Dassault}. The binary solvent mixture combinations were selected based on the miscibility table of Sigma Aldrich \cite{SigmaAldrich}, while the solutes were those used in our earlier work \cite{VermeireTL}. The data points were generated as part of another study on active learning, where the machine learning algorithm decided which data points need to be added to reduce the ensemble-based uncertainty of the model and automatically performs those COSMOtherm calculations. As such, more solutes were included with a higher molar mass, leading to a broader solvation free energy distribution. The database was compiled from all training and test sets of the former work, including those generated by the active learning algorithm as well as those generated from the comparison to randomly sampled data. This resulted in a database of 2.5 million data points from which we randomly sampled one million data points for the training procedure. More information on the generation of these datasets is given in the supporting information (SI).    

\subsubsection{Experimental databases for binary and ternary solvent mixtures}
Two datasets of experimental solvation free energies in mixed solvents were curated from data available in the Dortmund Data Bank (DDB) \cite{Dortmund}. 
\begin{figure*}[t]
    \centering
    \includegraphics[width=0.95\textwidth]{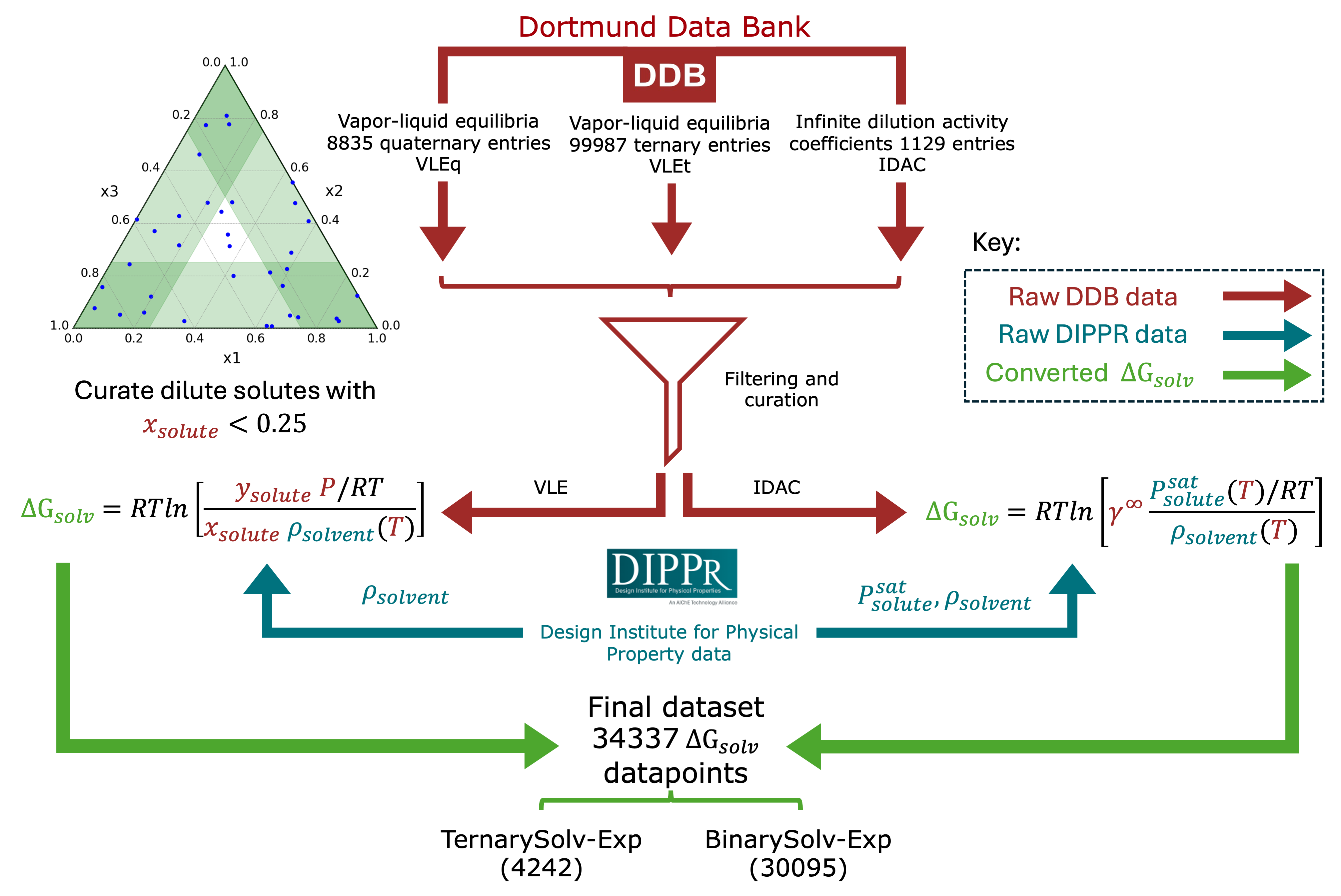}
    \caption{Diagram of the process used for experimental data curation and conversion.}
    \label{fig:diagram_simplified}
\end{figure*}
The theoretical approach was similar to an earlier work by Moine et al. \cite{Moine} to produce the CompSol dataset, which contains experimental solvation free energies in monosolvents (\textit{i.e} binary systems, where one component is infinitely dilute). The details of this theoretical approach are given in the SI. In our work, solvation energies were not measured directly but were calculated from experimental three- and four-component VLE data and binary solvent infinite dilution activity coefficient (IDAC) data. The VLE data includes temperatures $T$, pressures $P$, vapor phase compositions $y$, and liquid phase compositions $x$. This data was converted to infinite dilution solvation free energies of a solute $i$, $\Delta G^\infty_{i,\,solv}$, using \autoref{eq:VLE_gsolv_final_m}:
\begin{equation} \label{eq:VLE_gsolv_final_m}
 \Delta G^\infty_{i,\,solv}(T,P,\mathbf{x}) =  RT \ln\left[\frac{y_i}{x_i}\frac{P/RT}{\rho_{liq}(T,P,\mathbf{x})}\right] 
\end{equation}
where R is the universal gas constant and $\rho_{liq}$ is the solvent density. The density was estimated using correlations for pure components available in the DIPPR database \cite{dippr}.
Similarly, the IDAC data was converted using \autoref{eq:act_gsolv_final_m}:
\begin{equation} \label{eq:act_gsolv_final_m}
    \begin{aligned}
    \Delta G^\infty_{i,\,solv}(T,P,\mathbf{x}) 
    = RT \ln\left[\gamma^\infty_i(T, \mathbf{x})\frac{P_i^{sat}(T)/RT}{\rho_{liq}(T,P,\mathbf{x})}\right]
    \end{aligned}
\end{equation}
which also requires the vapor pressure of the solute, $P_i^{sat}$ that is available in DIPPR. A simplified overview of the data curation and conversion is shown in Figure \ref{fig:diagram_simplified}, where raw data are converted to 30095 solvation energies in binary solvents (BinarySolv-Exp) and 4242 solvation energies in ternary solvents (TernarySolv-Exp). The details of the curation process are provided in the SI.

\subsubsection{Dataset characteristics and comparative analysis} \label{DataCompar}
The characteristics of the experimental and synthetic datasets are compared and the impact of these characteristics on the training and testing procedure is discussed. An overview of the solvation free energy databases is given in \autoref{tab:databases}. On top of the solvation free energy, there are also values for the solvation enthalpy available for CombiSolv-Exp, CombiSolv-QM, and BinarySolv-QM; the details of these databases are discussed in the SI. 

\begin{table}[h!]
\caption{An overview of the databases used in the training and validation of the machine learning models. The abbreviation Combi is used for monosolvent data (from previous works \cite{VermeireTL}), Binary for binary solvents, and Ternary for ternary solvents. Additionally, Exp stands for experimental data while QM stands for quantum chemistry derived data.}
\label{tab:databases}
\footnotesize
\begin{tabular}{|l|l|l|l|l|}
\hline
\rowcolor[HTML]{EFEFEF} 
\textbf{Database}          & \textbf{Data points} & \textbf{Unique solvents} & \textbf{Unique solutes} & \textbf{Temp}      \\ \hline
CombiSolv-Exp     & 8,780       & 275             & 1,307          & $298 K$   \\ \hline
BinarySolv-Exp    & 30,095      & 204             & 210            & $T_{exp}$ \\ \hline
TernarySolv-Exp   & 4,242       & 66              & 62             & $T_{exp}$ \\ \hline
CombiSolv-QM      & 1,000,000   & 284             & 10,836         & $298 K$   \\ \hline
BinarySolv-QM     & 1,000,000   & 32              & 10,960         & $298 K$   \\ \hline
\end{tabular}
\end{table}

The two QM databases differ mainly in the number of unique solvents. BinarySolv-QM contains significantly fewer unique solvents than CombiSolv-QM, which is due to the fact that BinarySolv-QM is limited to pairs of solvents that are known to be miscible. An equal number of data points were randomly sampled from each of the QM databases for the creation of the training dataset to have both a diverse set of unique solvent molecules and mixed solvents included.

When analyzing experimental databases, a clear trend emerges: the diversity of unique solvents and solutes diminishes as the number of components in the mixture increases. This pattern is generally anticipated when comparing the data available for pure substances and mixtures \cite{fabian}. So although BinarySolv-Exp contains significantly more data points compared to CombiSolv-Exp, this is mainly due to a given binary solvent system being repeated for different compositions, whereas CombiSolv-Exp only contains monosolvents.

Experimental datasets for thermochemical properties are known to be noisy, especially if curated from a wide variety of sources as is the case for the DDB \cite{Dortmund}. 
Experimental noise, also called aleatoric uncertainty, can have a large influence on the perceived performance of machine learning models. When different model architectures are compared on a noisy dataset, they can seemingly have the same performance (equal to the aleatoric limit) even if one is actually significantly better then the other \cite{Heid}. The COSMOtherm datasets contain low aleatoric uncertainty. For this reason, the later proposed ML architectures were compared based on the the synthetic datasets.

\begin{figure}[h!]
\centering
\begin{subfigure}{.5\textwidth}
  \centering
  \includegraphics[width=\linewidth]{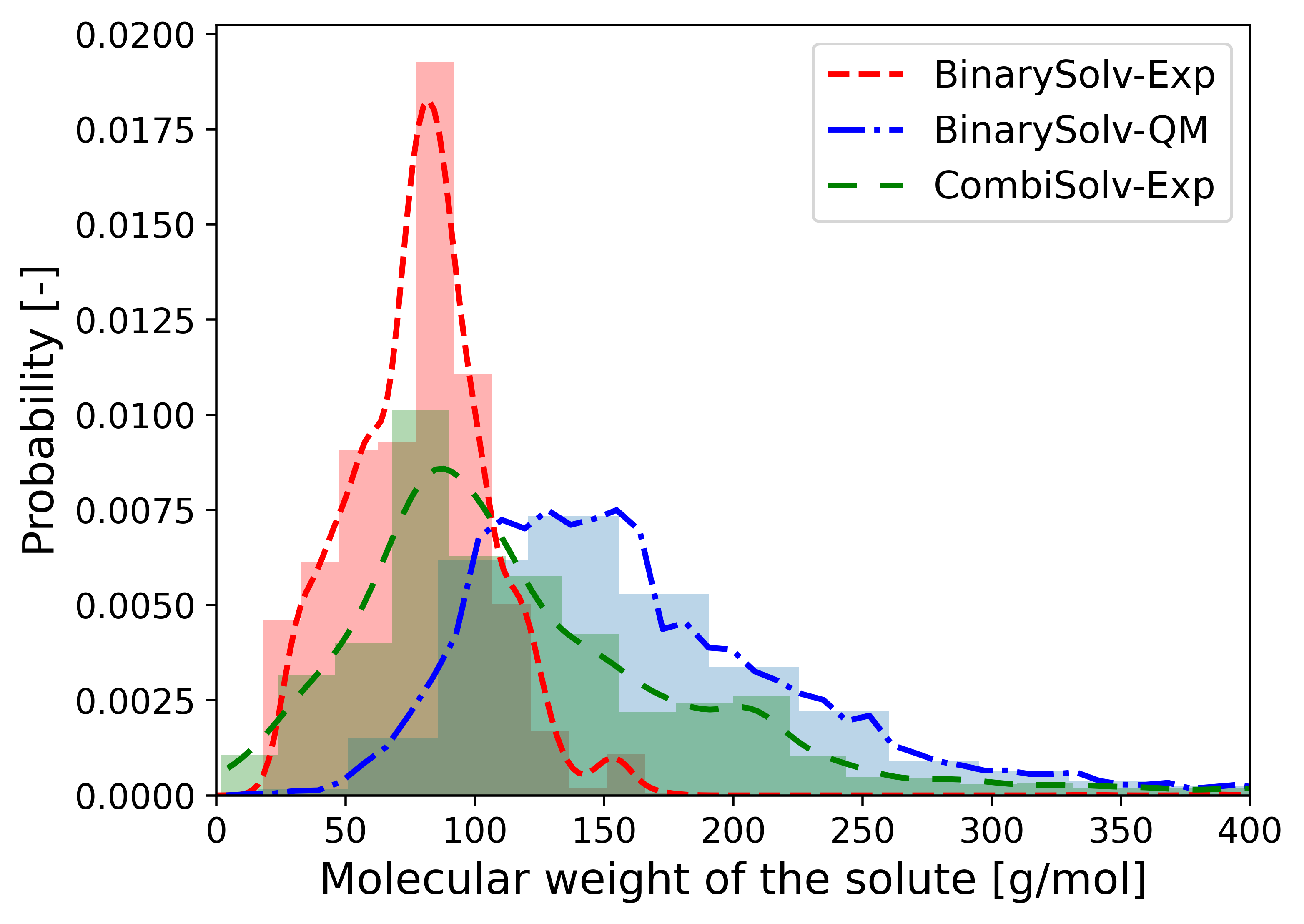}
  \caption{}
  \label{fig:multi_qm_vle_molweight}
\end{subfigure}%
\begin{subfigure}{.5\textwidth}
  \centering
  \includegraphics[width=0.95\linewidth]{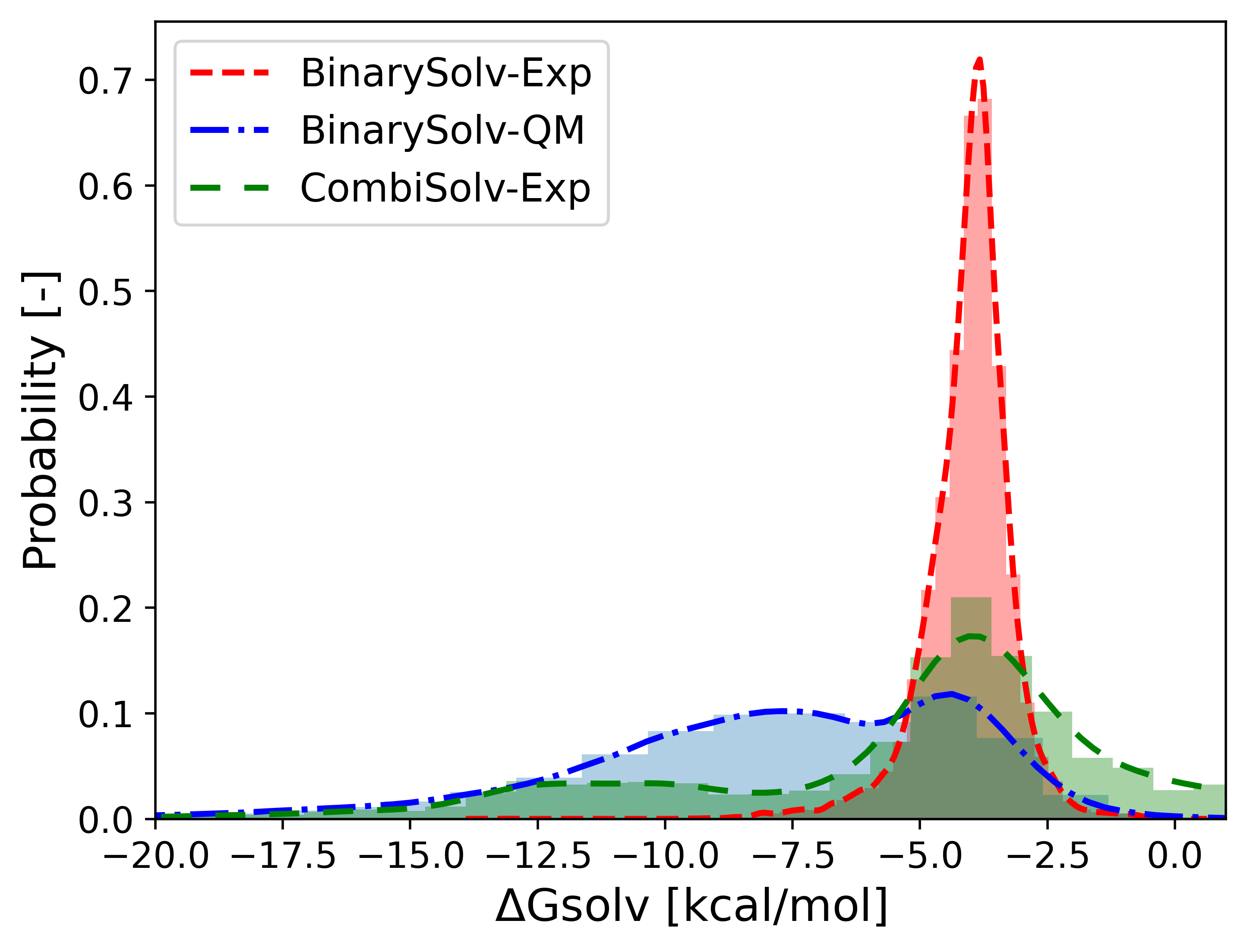}
  \caption{}
  \label{fig:multi_pure_qm_dGsolv}
\end{subfigure}
\caption{(a) Molecular weight and (b) solvation free energy distributions of the experimental (red) and quantum chemical (blue) datasets. }
\label{fig:distributions1}
\end{figure}

In \autoref{fig:distributions1}, we present the distributions of molecular weight and solvation free energy for the BinarySolv-Exp, BinarySolv-QM, and CombiSolv-Exp (monosolvent) databases. For clarity of the figures, the TernarySolv-Exp and CombiSolv-QM databases are not included in these plots. A detailed comparison between TernarySolv-Exp and BinarySolv-Exp distributions can be found in the SI, where it is evident that TernarySolv-Exp has even narrower distributions due to the limited experimental data available for quaternary systems. The BinarySolv-QM and CombiSolv-QM databases show similar distributions, as both contain the same solutes.

From \autoref{fig:distributions1}(a), it can be seen that the distribution of the molecular weight of the solutes in the BinarySolv-Exp database is much smaller as compared to the BinarySolv-QM database, while the distribution for the CombiSolv-Exp database falls between those of the BinarySolv-Exp and BinarySolv-QM databases. 
For the BinarySolv-Exp database, the majority of the molecular weights of solutes are under 100 g/mol, which also results in smaller magnitudes for the solvation free energies. This imbalance towards smaller solutes is due to deriving the data primarily from VLE measurement which is limited to species that have measurable vapor pressures at conditions where common solvents are liquids. For the dataset calculated with COSMOtherm, on the other hand, there is an opportunity to cover a broader chemical space by selecting heavier molecules for the calculations. Large molecules usually have more negative $\Delta H_{solv}$ and so more negative $\Delta G_{solv}$ at moderate temperatures, see \autoref{fig:multi_pure_qm_dGsolv}.

When a dataset is considered for training a machine learning model, it is important for the robustness of the model that the dataset covers of a broad and diverse molecular space. 
Otherwise, the generalizability and applicability of the model will be compromised. The COSMOtherm datasets make it possible to train models for a wide applicability range. As the CombiSolv-Exp distribution has a wider spread of solutes compared to BinarySolv-Exp and TernarySolv-Exp, this database was used for fine-tuning the machine learning models.
The mixed solvent experimental datasets were only used for testing the performance of the models because of their limited molecular diversity. 

\begin{figure}[h!]
    \centering
    \includegraphics[width=7.5cm]{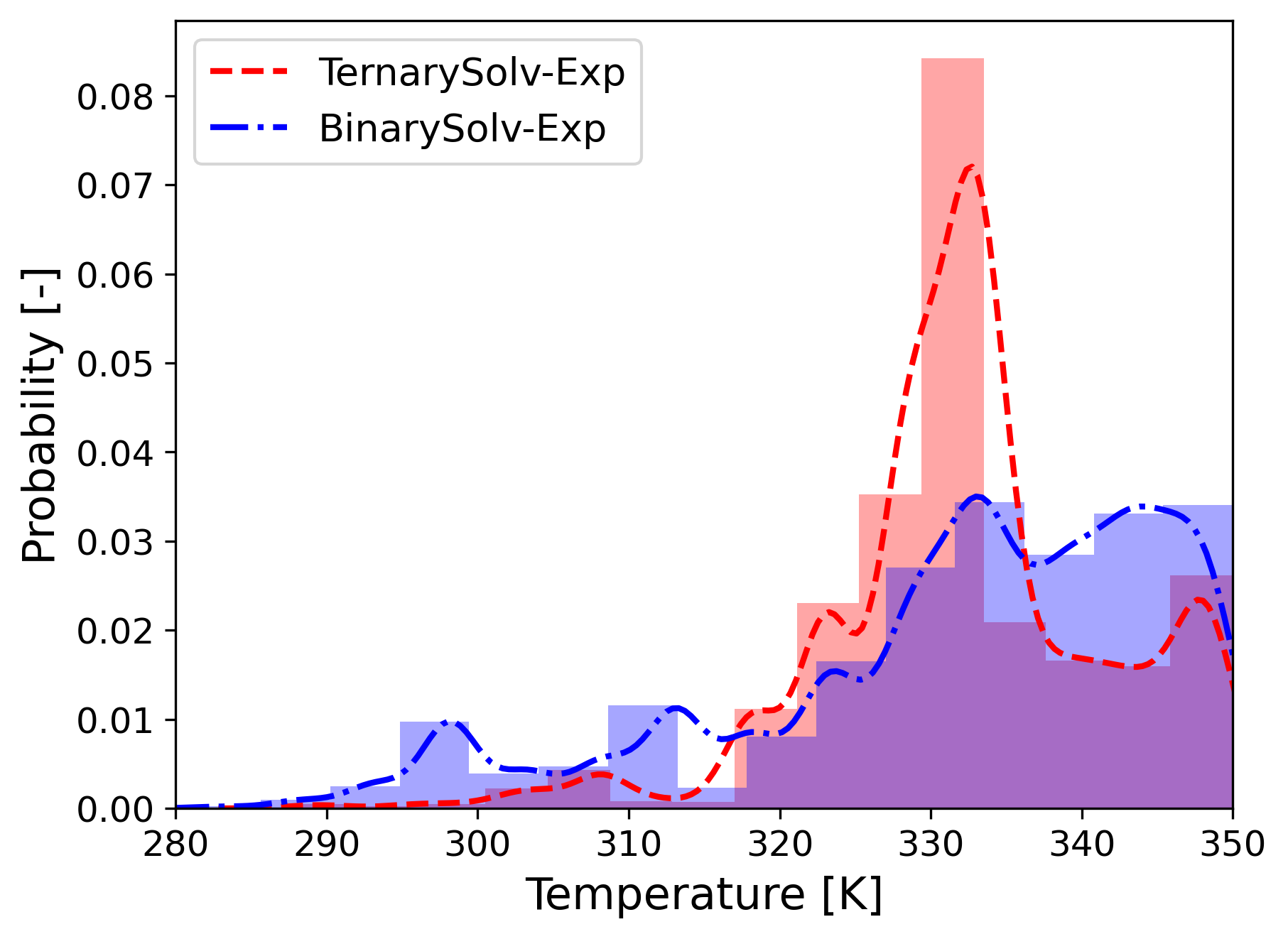}
    \caption{Temperature distribution of the experimental datasets.}
    \label{fig:temp_distribution}
\end{figure}

The datasets BinarySolv-Exp and TernarySolv-Exp differ from the other datasets in this work with respect to temperature. The experimental mixed solvent data contain a temperature range from $280K$ to $350K$ as shown \autoref{fig:temp_distribution}, where all the other datasets are reported at $298K$.
Therefore, the temperature dependence of solvation free energy for the prediction on these datasets was taken into account as discussed with the training procedure.

\subsection{Machine learning methods}
\subsubsection{Directed message passing neural network}

Graph neural networks take graph data structures as input and learn molecular embeddings (\textit{i.e.}, molecular fingerprints) based on their topology. Put differently, based on a graph representation of a molecule, a learned molecular fingerprint is created. Chemprop \cite{Chemprop} is a Python repository for molecular property prediction using a type of GNN called a directed message passing neural network (D-MPNN). The repository used, SolProp, is based on the same D-MPNN framework and focuses specifically on solvation property prediction \cite{VermeireSolu}. D-MPNNs create molecular embeddings in three phases: (1) the set up phase, where hidden directed-bond vectors are initialized from atom and bond feature vectors, (2) the message passing phase, where the hidden directed-bond vectors are updated based on the neighbouring ones, and (3) the readout or pooling phase, where the directed bonds are read out to the atoms they point to and subsequently the atom feature vectors are pooled to a molecular fingerprint. 
All the molecular fingerprints used in the results section were created by D-MPNNs as implemented in Chemprop, using the initial atom and bond features as implemented in SolProp. These features are listed in the SI. For more details regarding the D-MPNN framework, the reader is referred to dedicated work \cite{Chemprop, VermeireTL}. 

\subsubsection{SolProp-mix and alternative architectures for mixture representation} \label{architectures_method}

\begin{figure}[h]
    \centering
    \includegraphics[width=\textwidth]{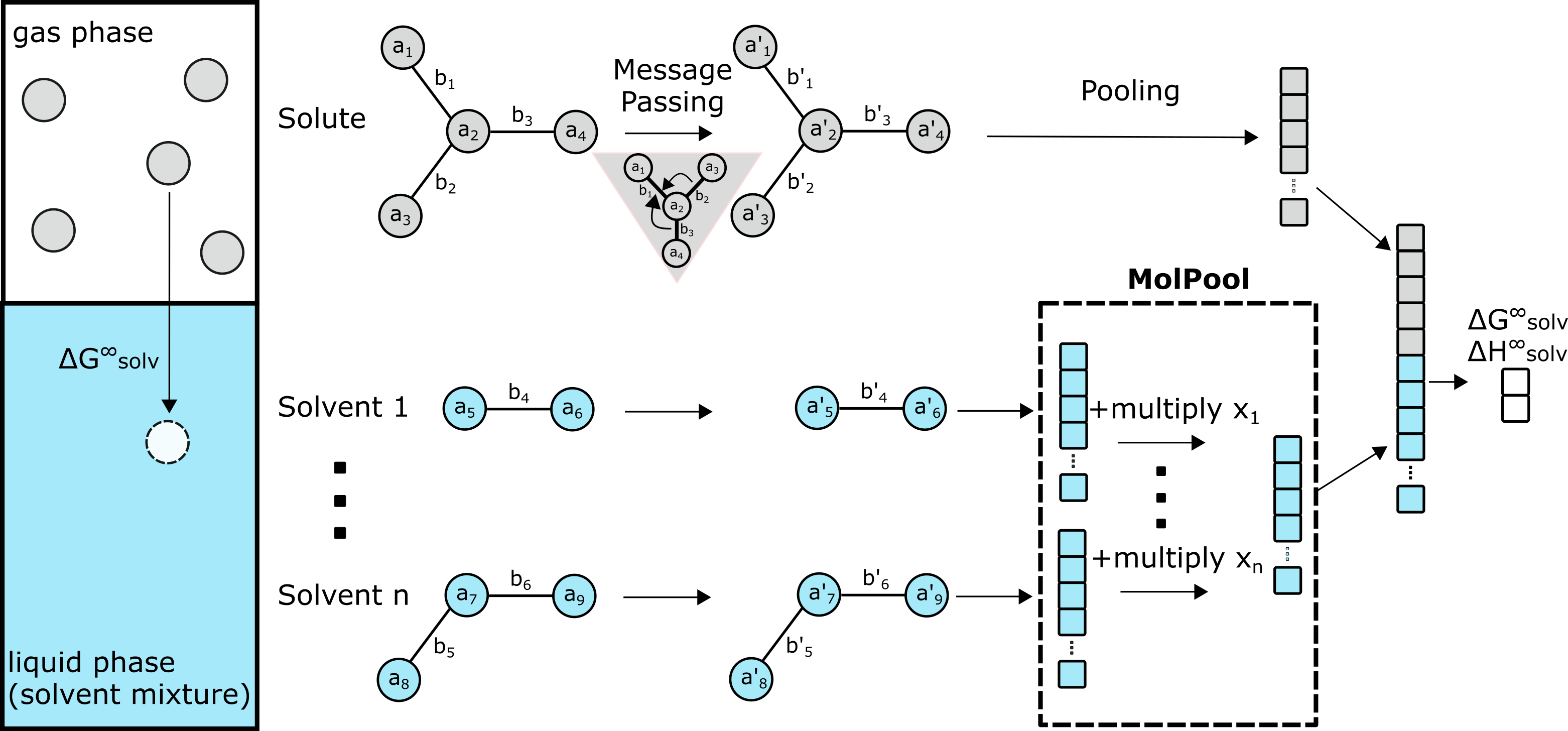}
    \caption{he left box gives a representation of the solvation free energy process. To the right of this box the SolProp-mix architecture is shown including the pooling function MolPool in the dashed box.}
    \label{fig:mpn_pooling}
\end{figure}
In our proposed architecture, SolProp-mix, for solvation property prediction (see \autoref{fig:mpn_pooling}), two separate D-MPNNs are used: one for the solute and one for the solvents in the mixture. The pooling of the components in the solvent mixture using our proposed function, MolPool, is the novelty in this work compared to previous works. We recently applied this method also to the analysis of fuel properties, demonstrating its flexibility \cite{Leenhouts}.
MolPool constructs the embedding for a mixture by averaging the embedding of each component in the mixture, weighted by its mole fraction. This is shown in the dashed box in \autoref{fig:mpn_pooling}. 
The mixture embeddings produced by this pooling function have two important properties that make it permutation invariant: (1) its length is constant regardless of the number of components in the mixture and (2) it does not depend on the input order of the solvents.
This solvent (mixture) embedding is finally concatenated to the molecular fingerprint of the solute and send through a multi-layer perceptron (MLP). The weights of each D-MPNN are learned at the same time as the weights of the MLP. That is, the whole model is trained end-to-end. 

\begin{figure}[h]
    \centering
    \includegraphics[scale=0.85]{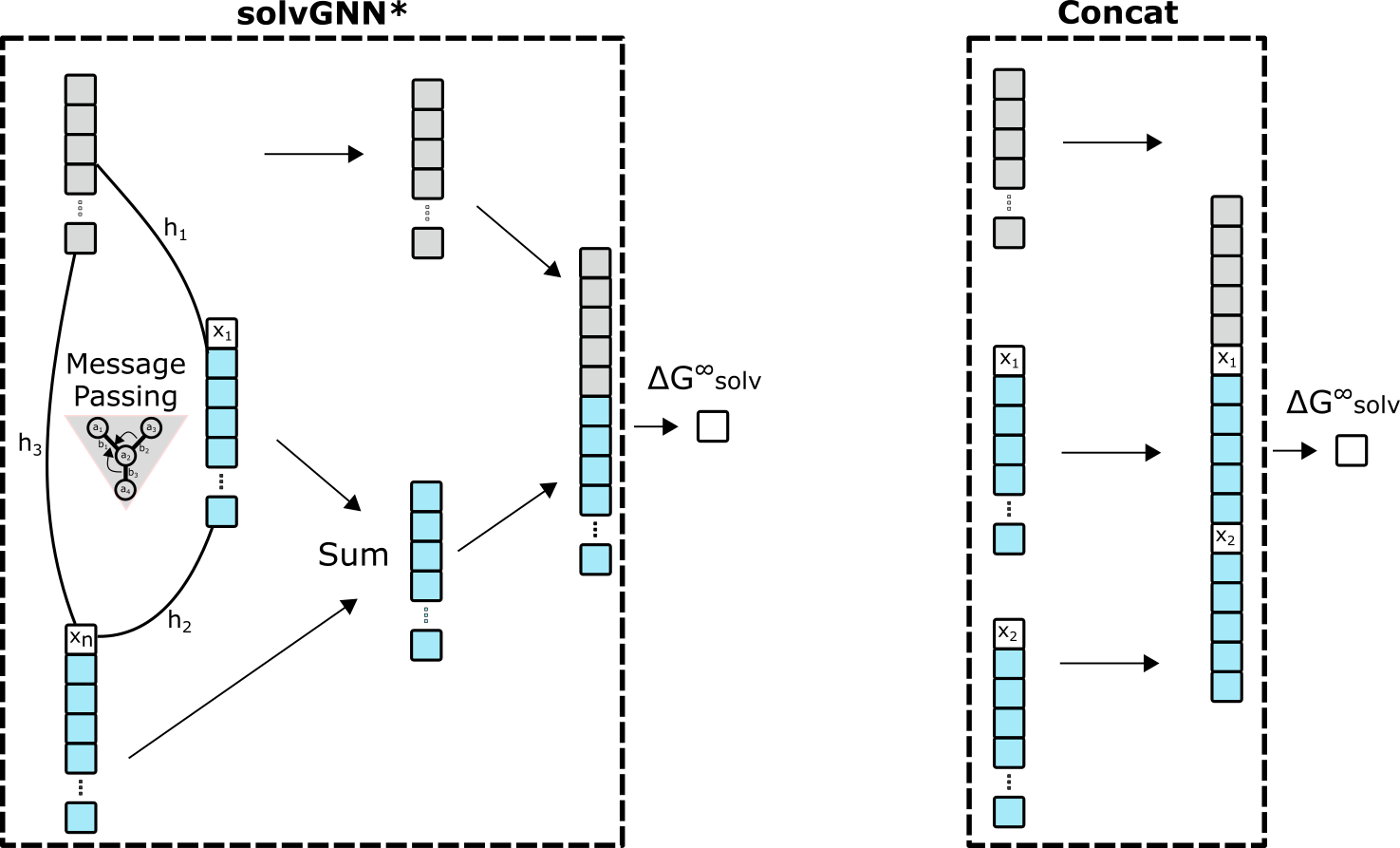}
    \caption{Pooling function for the implementation of solvGNN* and Concat architectures. In the implementation, these replace the dashed box in the SolProp-mix architecture.}
    \label{fig:concat_solv}
\end{figure}

The SolProp-mix architecture (\autoref{fig:mpn_pooling}) was compared to the two alternative architectures for mixture representations. First, a modified version of the solvGNN architecture, as published by Qin et al. \cite{Qin}, was used (called solvGNN* in this work). 
For solvGNN*, we used a D-MPNN as implemented in SolProp instead of the original solvGNN graph convolution to generate the molecule embeddings. The SolvGNN layer then generates a mixture graph of the molecule embeddings, where the bonds are defined based on Lipinksi hydrogen bond information, see \autoref{fig:concat_solv} on the left.
A modification to the original solvGNN layer was required, since solvGNN \cite{Qin} was created for activity coefficients, while here the target is solvation free energy. For activity coefficients, predictions are typically made for every component in the mixture, where for infinite dilute solvation free energies the interest only lies in the energy of the dissolving solute molecule.
Therefore, the permutation equivariant concatenation from the original solvGNN paper \cite{Qin} was replaced by the permutation invariant summation.hIn solvGNN, the mole fractions of the solvents are concatenated to the molecular fingerprints. The solute fingerprint is then concatenated to the mixture embedding. 

A second comparison was made to a concatenation architecture, as there are several examples in the literature that used such architectures to represent molecular mixtures \cite{Winter, Medina, Winter2}.
In this architecture, the individual component embeddings are concatenated along with their mole fractions, as shown in \autoref{fig:concat_solv} (right). This concatenated mixture embedding is the input for an MLP. This architecture is neither permutation invariant or extendable to an arbitrary number of solvents. The MLP of a model expects a fixed-length embedding, but the embedding for binary solvent data, for example, will be twice the length of the embedding for monosolvent data. To account for this, a workaround was employed which involved using an MLP configured to accept the longest mixture embedding in the dataset, with shorter embeddings padded with zeros as needed. Different approaches can be considered for incorporating mole fractions into this architecture to improve its generalization capabilities, though ensuring permutation invariance remains a challenge.

\subsubsection{Training procedure}
Model training occurred in two stages. During the first stage, synthetic data from the CombiSolv-QM and BinarySolv-QM databases was used to train and compare the different ML architectures proposed in the last section. During the second stage, the pre-trained model weights of the SolProp-mix and Concat architectures were further optimized (\textit{i.e.} fine-tuned) on the CombiSolv-Exp database. For this fine-tuning process, any data points with solute molecules present in both CombiSolv-Exp and the test datasets (BinarySolv-Exp and TernarySolv-Exp) were removed from the CombiSolv-Exp database. Finally, their performance on the BinarySolv-Exp and TernarySolv-Exp databases was compared to COSMOtherm calculations. 

All models were trained as multitask models for infinite dilution solvation free energy and enthalpy at $298 K$. This allows the models to account for moderate temperature dependence, using \autoref{eq:gsolv_T}. In this equation $\Delta G^\infty_{solv,\,298\,K}$ is the solvation free energy, $\Delta H^\infty_{solv,\,298\,K}$ is the solvation enthalpy, and $T$ is the temperature. This equation is based on the van't Hoff equation and assumes that solvation enthalpy does not depend on temperature. This assumption holds the best for temperatures near $298 K$. In this work, we extrapolate to a maximum temperature of $350 K$, and found that prediction errors did not increase at elevated temperatures below $350 K$. See the SI for more details. This temperature extrapolation was applied to test the models on the BinarySolv-Exp and TernarySolv-Exp databases.

\begin{equation} \label{eq:gsolv_T}
    \Delta G^\infty_{solv,\,T} = T\left(\frac{\Delta G^\infty_{solv,\,298\,K}}{298\,K} - \Delta H^\infty_{solv,\,298\,K}\left(\frac{1}{298\,K} - \frac{1}{T}\right)\right)
\end{equation}

The first stage of training was performed on an increasing number of data points to investigate the effect of dataset size on the performance of the three architectures. Data was randomly sampled equally from the CombiSolv-QM and BinarySolv-QM databases ($10^3$, $10^4$, $10^5$, and $10^6$ data points). More details on the training procedure, the hyperparameters and features are given in the SI.

The procedure for splitting data merits special consideration for solvation properties, as a solute may appear in both the training and test sets if it is paired with multiple solvents causing data leakage. Because the main focus of this research was to predict the solvation free energies of new active pharmaceutical ingredients (solutes) in known solvents, we ensured that all data points involving a given solute appeared in only one of the training, validation, or test sets. 

We used ten fold cross-validation to split the data into datasets for training, validation, and testing. Each data point appeared in a test set only once and ten models were trained. Thus in the results section, parity plots show the test set prediction of all ten test sets joined together while the overall model performance reported is the average performance of the ten individual models. The repeatability and robustness of an architecture can be judged from the standard deviation of the performance of the ten individual models.

The second stage of training involved fine-tuning the models using the CombiSolv-Exp database. For the fine-tuning step, the parameters stemming from the training on the QM databases were used for the initialization of the network. During the fine-tuning, all the network weights were unfrozen, \textit{i.e.} were allowed to optimize. A ten fold cross-validation was used to split the data into training, and validation sets. The experimentally fine-tuned models were tested on the BinarySolv-Exp and TernarySolv-Exp databases. These databases are limited in diversity, so this test was not a measure of the absolute performance, but was used as a comparison to COSMOtherm calculations. The solutes in the BinarySolv-Exp and TernarySolv-Exp database do not overlap with the solutes in CombiSolv-Exp, but there is some overlap with the solutes in the synthetic COSMO-RS datasets. This should not affect the relative performance of our models and COSMOtherm given neither had access to experimental data points for these solutes. 

\section{Results \& Discussion}
\subsection{Architecture comparison}
\subsubsection{Learning curves for the proposed architectures}
\begin{figure}[h]
    \centering
    \includegraphics[scale=0.60]{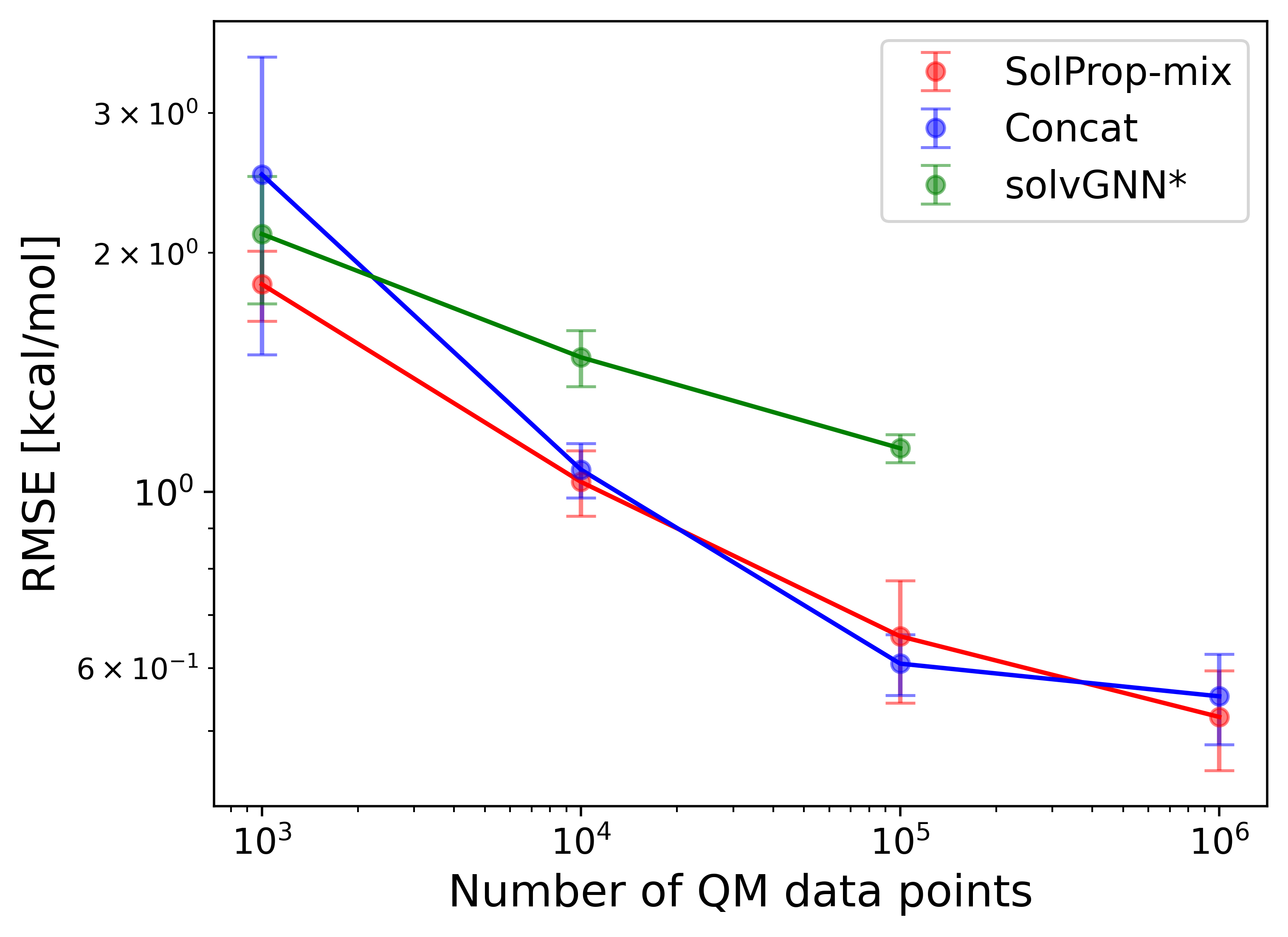}
    \caption{The average RMSE on 10\% of the merged CombiSolv-QM and BinarySolv-QM databases versus the dataset size performance for the three architectures. The error is shown over ten fold cross-validation. The error bars indicate the standard deviation over these folds.}
    \label{fig:architectures}
\end{figure}

To assess the predictive performance of various architectures for solvation free energies in solvent mixtures, we analyzed three distinct architectures as described in Section \ref{architectures_method}. The results of these comparisons for different dataset sizes are presented as a learning curve in \autoref{fig:architectures}, which presents the average root mean squared error (RMSE) for the merged QM databases, BinarySolv-QM and CombiSolv-QM. The test set for each fold is 10\% of the merged QM database, based on solute splitting. 
Error bars in the plot indicate the standard deviation of errors across the different test sets. A smaller standard deviation suggests a more consistent performance. 

As expected, the performance of all architectures improved with an increasing number of data points. 
The RMSE of the solvGNN* architecture was higher compared to SolProp-mix from $10^3$ up to $10^5$ data points.
The Concat architecture initially showed higher average error and variability at $10^3$ data points but converged with SolProp-mix’s error levels around $10^4$ data points. The problem with the Concat architectur, however, is its invariance to the solvent order.
To illustrate the difference in permutation invariance between the Concat and SolProp-mix architectures, \newline predictions were made for paracetamol in a 50\% water and 50\% acetone mixture by the models trained on $10^6$ QM data points. When the solvent order was reversed, the Concat architecture produced predictions of $-14.5 \ kcal/mol$ and $-16.4 \ kcal/mol$ for the same mixture, while SolProp-mix, by applying MolPool, consistently predicted $-16.5 \ kcal/mol$ in both cases.

The Concat and SolProp-mix architecture required a similar amount of training time, since they have the same number of neural network layers and a similar number of neural network weights.
The solvGNN* architecture, on the contrary, required significantly more training time and memory compared to Concat and SolProp-mix. 
Where the training of the other architectures ran on 64 GiB RAM for the $10^6$ data point, the solvGNN* model's memory demands exceeded 256 GiB of RAM, making it infeasible to process the largest dataset, highlighting potential inefficiencies in the model's design. This data point had to be left out of the learning curve. 

\begin{table}[]
\caption{Overview of the training details for the different architectures used in this work.}
\label{tab:architecture_data}
\footnotesize
\begin{tabular}{|l|l|l|l|}
\hline
\rowcolor[HTML]{EFEFEF} 
\textbf{Architecture} & \textbf{No. parameters} & \textbf{Invariance} & \textbf{Memory for $10^6$ data point} \\ \hline
SolProp-mix      & 902k                 & Yes                   & $64$ GiB RAM                    \\ \hline
Concat       & 1.00M                 & No                   & $64$ GiB RAM                    \\ \hline
solvGNN*      & 1.50M                & Yes                   & $>256$ GiB RAM                  \\ \hline
\end{tabular}
\end{table}

SolProp-mix was compared to the solvGNN* implementation to investigate whether the inclusion of molecular interactions improves the performance.
In the solvGNN* architecture, the molecular interactions between components in a mixture are not learned, but defined based on Lipinski hydrogen bonds or "hypothetical hydrogen bonds" as mentioned in the work of Qin et al. \cite{Qin}. However, interactions between molecules in a mixture are often more complex than Lipinski hydrogen bonds. By implying these interactions, the model could learn a wrong or incomplete interaction behaviour and these errors then propagate through the network. In any event, the increase in computational resources for solvGNN*, does not result in a performance increase for this application. Overall, the SolProp-mix model demonstrated the best balance of accuracy, consistency, and efficiency, making it suitable for applications that require scalable and reliable predictions.

\subsubsection{Generalization from monosolvent to binary solvents}
\begin{figure}[h!]
\centering
\begin{subfigure}{.5\textwidth}
  \centering
  \includegraphics[width=\linewidth]{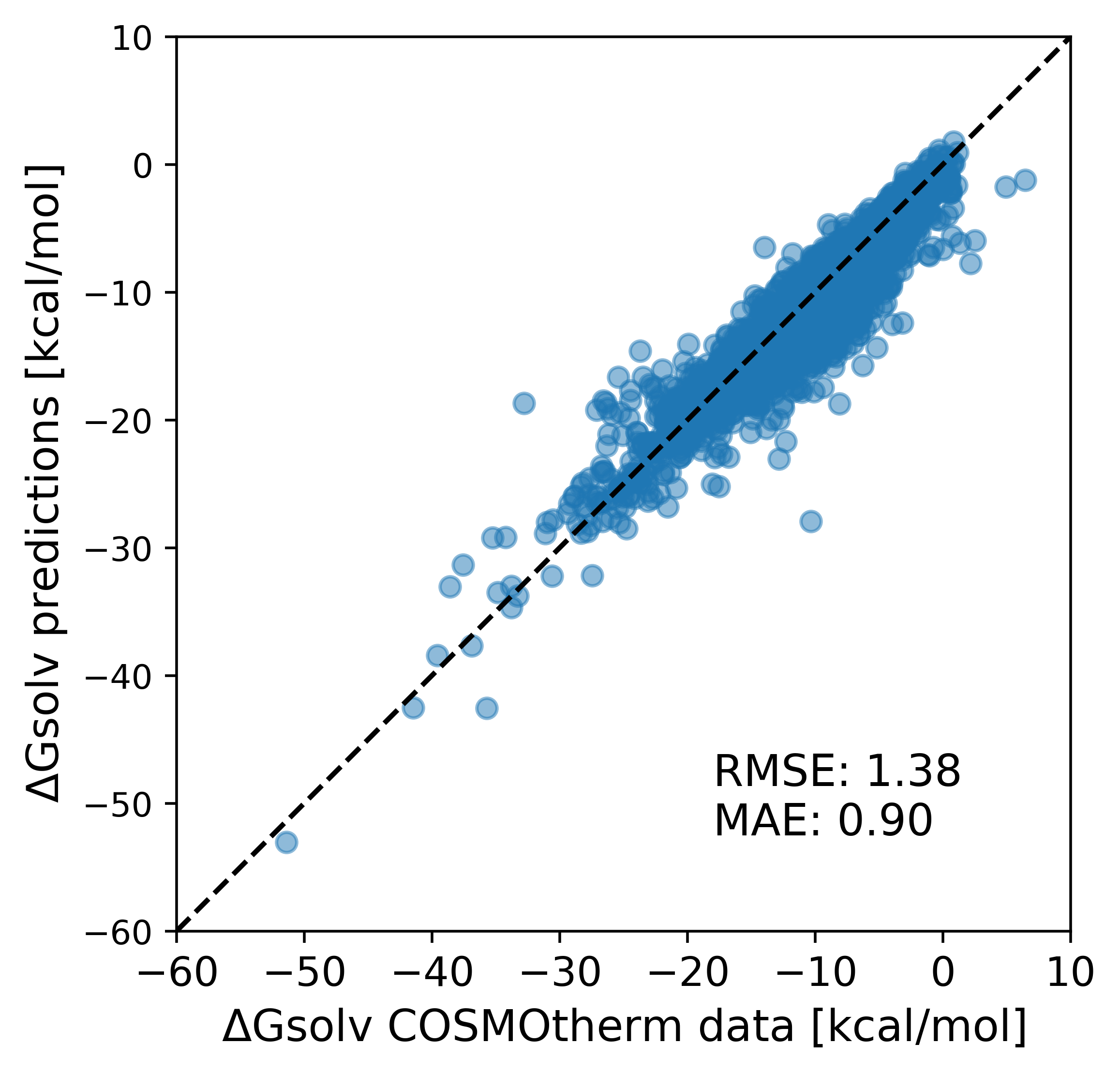}
  \caption{Concat architecture}
  \label{fig:sub1}
\end{subfigure}%
\begin{subfigure}{.5\textwidth}
  \centering
  \includegraphics[width=\linewidth]{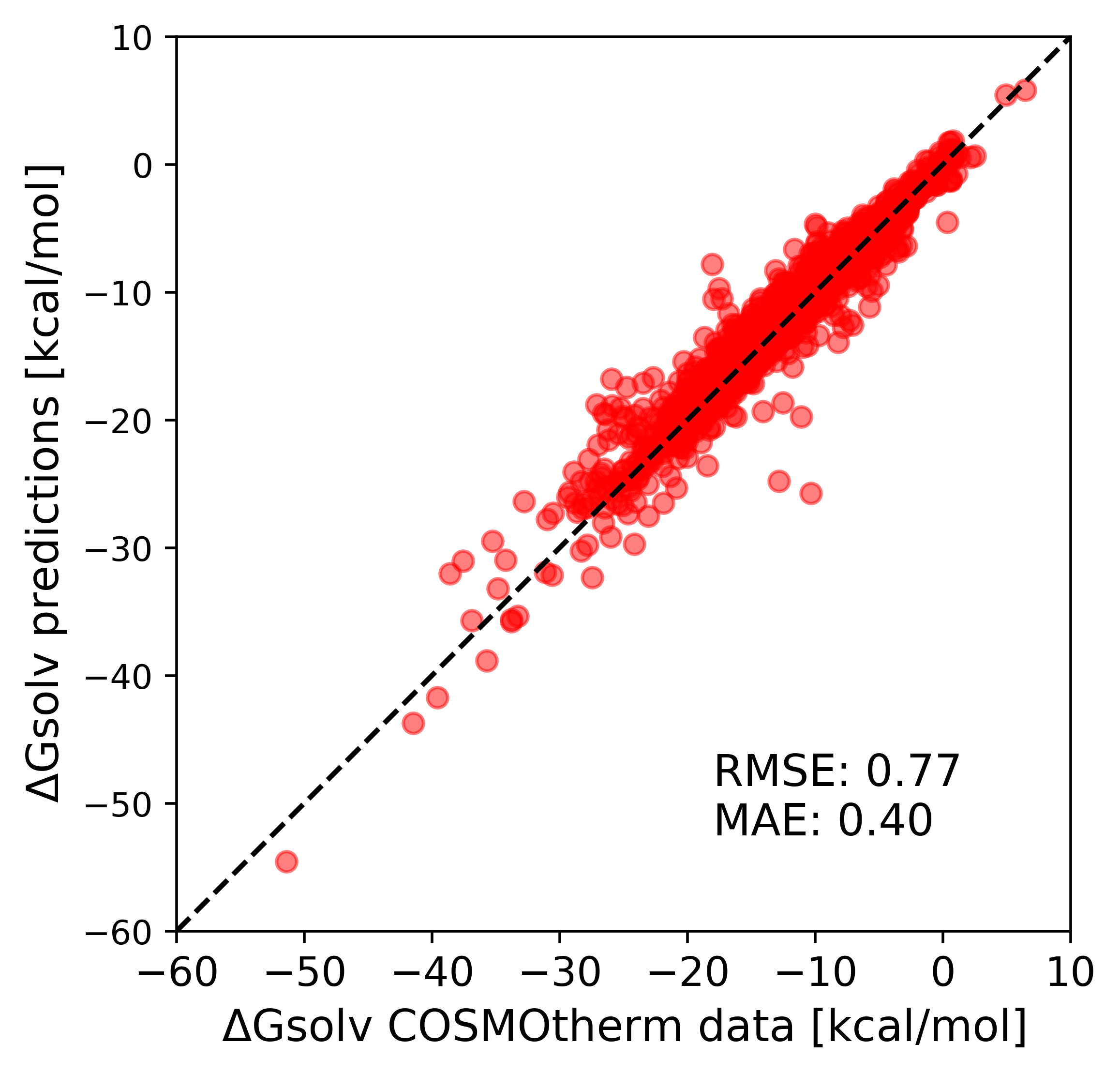}
  \caption{SolProp-mix architecture}
  \label{fig:sub2}
\end{subfigure}
\caption{The predictions on solvent mixture data for the SolProp-mix and Concat architectures trained on monosolvents. The results are made by model ensembles trained and validated on ten fold cross-validation.}
\label{fig:test}
\end{figure}

A useful model architecture for molecular mixtures should include a pooling function that not only accepts an arbitrary number of components, but also generalize information from one number of components to another. To test this explicitly, we applied ten fold cross validation to $10^5$ monosolvent data points from CombiSolv-QM and trained models using both the Concat and SolProp-mix architectures. Each ensemble of models was then tested on $10^4$ binary solvent mixtures from BinarySolv-QM. The results are shown in \autoref{fig:test}. 

As expected, the accuracy for both the Concat and the SolProp-mix architectures decreased compared to the previous plot, \autoref{fig:architectures}, where the average RMSE for training on $10^5$ sampled binary solvents and monosolvent was $0.61 \ kcal/mol$ for Concat and $0.66 \ kcal/mol$ for SolProp-mix. However, when training on monosolvent data and testing on binary solvent mixtures, SolProp-mix performed significantly better compared to the Concat architecture, improving the RMSE from $1.38$ to $0.78 \ kcal/mol$. It is important to note that the Concat architecture has a much higher performance when trained on a combination of monosolvents and binary solvents as shown in \autoref{fig:architectures}, while the model accuracy quickly drops when only monosolvent data is used. 
This is likely due to some of the model weights of the Concat architecture not being updated during training as the monosolvent representation was padded with zeros to account for the embedding being shorter than for binary sovlents. 

Remarkably, for SolProp-mix, training exclusively on monosolvent data resulted in only a minor decrease in performance compared to training on both monosolvent and binary solvent data. This architecture is able to learn properties of mixtures of molecules by only training on monosolvent data.  
The fact that pure or binary data can be used for the prediction of thermodynamic properties in more complex mixtures has been demonstrated before. For example, binary mixture data can be used for Gibbs excess predictions in ternary mixtures \cite{Prausnitz, Abrams_Prausnitz_1975}. Also for the Hansen solubility parameters, interpolation of the distance metrics of individual solvents was used in the past for the prediction of solubility in solvent mixtures \cite{Hansen_2012}. The novelty of this work is that through the use of graph neural networks, the MolPool function makes it possible to predict properties in more complex mixtures of unseen compounds, where conventional methods always rely on the availability of pure or binary component data
\subsection{Model performance on experimental datasets}
\subsubsection{Effect of transfer learning on predictions for BinarySolv-Exp}essfully generalizes its performance to solvent mixtures.QM data for binary solvents In this section, it was hypothesized that fine-tuning a model on monosolvent experimental data would improve performance on predicting solvation free energy in experimental binary solvents. This would be promising, because the database comparison showed that the data in monosolvents is more abundant and diverse compared to the data in binary and ternary solvents. Below, the results of the transfer learning procedure, as described in the training procedure section, are shown for the prediction of experimental solvation free energies in the new dataset for binary solvents \textit{i.e.}, BinarySolv-Exp. Comparisons are made between the Concat and SolProp-mix architectures, while COSMOtherm calculations are used as a benchmark. Table \ref{tab:performance} shows the mean absolute error (MAE), RMSE, and standard deviation of the model ensemble for COSMOtherm calculations and graph neural networks trained on different subsets of the databases.

\begin{table}[h!]
\caption{The performance of different architectures and benchmarks tested on the BinarySolv-Exp database. "QM" in the training data column refers to the merged CombiSolv-QM and BinarySolv-QM database. $\sigma$ refers to the average standard deviation over the test data for the model ensemble.}\label{tab:performance}
\footnotesize
\begin{tabular}{|l|l|l|l|l|}
\hline
Model       & Training data       & MAE(kcal/mol) & RMSE(kcal/mol)  & $\sigma$(kcal/mol)  \\ \hline
COSMOtherm  & -                   & 0.36          & 0.50            & -                   \\ \hline
SolProp-mix & QM                  & 0.32          & 0.46            & 0.10                \\ \hline
SolProp-mix & CombiSolv-Exp       & 0.38          & 0.59            & 0.14                \\ \hline
SolProp-mix & QM \& CombiSolv-Exp & \textbf{0.29} & \textbf{0.45}   & 0.073               \\ \hline
Concat      & QM                  & 0.33          & 0.49            & 0.13                \\ \hline
Concat      & QM \& CombiSolv-Exp & 0.37          & 0.57            & 0.096               \\ \hline
\end{tabular}
\end{table} D
 general, the reported errors on by the machine learning models this experimental database (RMSE $\pm 0.5\ kcal/mol$) a, see Table \ref{tab:performance}re smaller compared to the predictions for the QM datasets (RMSE $\pm 0.7\ kcal/mol$), while the opposite could be expected because of the experimental nature of this data.
However, this can be explained by the nature of the molecules in the VLE dataset. The molecules obtained from VLE measurements were significantly smaller in terms of molecular weight, and therefore also in solvation free energy magnitude. This means the magnitude in the error of the predictions can also be expected to be smaller. 

When comparing the performance of the pre-trained and fine-tuned models on the entire BinarySolv-Exp database, it appears that transfer learning only had a minor impact. For the SolProp-mix models, the model trained on only QM data had an RMSE of $0.49\ kcal/mol$ while the fine-tuned model had an RMSE of $0.46\ kcal/mol$. In comparison, for the Concat models, the model trained on only QM data had an RMSE of $0.49\ kcal/mol$ and the fine-tuned model had an RMSE of $0.50\ kcal/mol$. 
It should be noted that even though small differences can be observed between the different models, if the standard deviation from the model ensembles is account for it could be said that all models perform equally well. It is expected that performance shown in Table \ref{tab:performance} is at the aleatoric limit of the experimental dataset. However, it is difficult to quantify the uncertainty of the experimental measurements because of the scarcity of repeated experimental measurements. When the models reach the aleatoric uncertainty of the test set, it is no longer possible to differentiate the performance between the different model architectures.

While we observe a minor effect of transfer learning on the performance on the whole BinarySolv-Exp database, when the performance on systems with and without water as component in the solvent mixture was viewed separately, a more distinct effect was observed. The results for systems with water are shown in \autoref{tab:performance_water}, and the results for systems without water are shown in \autoref{tab:performance_no_water}. The SolProp-mix model trained on only QM data performed similar to COSMOtherm on the aqueous data. It also performed significantly better than the SolProp-mix model trained on only experimental monosolvent data. Fine-tuning the model on the experimental monosolvent data also caused the model to perform worse. When the same analysis is applied to the non-aqueous data, the fine-tuned model performs slightly better (roughly 20\%) than the pre-trained model even though a SolProp-mix model trained on only experimental monosolvent data performed slightly worse. 

\begin{table}[h!]
\caption{The performance of different architectures and benchmarks on the roughly 7500 data points in the BinarySolv-Exp database that include water as one of the solvents. "QM" in the training data column refers to the merged CombiSolv-QM and BinarySolv-QM database. $\sigma$ refers to the average standard deviation over the test data for the model ensemble.} 
\label{tab:performance_water}
\footnotesize
\begin{tabular}{|l|l|l|l|l|}
\hline
Model       & Training data       & MAE(kcal/mol) & RMSE(kcal/mol) & $\sigma$(kcal/mol)  \\ \hline 
COSMOtherm  & -                   & 0.40          & 0.58           & -                   \\ \hline 
SolProp-mix & QM                  & \textbf{0.37} & \textbf{0.57}  & 0.14                \\ \hline 
SolProp-mix & CombiSolv-Exp       & 0.65          & 0.93           & 0.16                \\ \hline 
SolProp-mix & QM \& CombiSolv-Exp & 0.41          & 0.62           & 0.086               \\ \hline 
Concat      & QM                  & 0.43          & 0.66           & 0.19                \\ \hline 
Concat      & QM \& CombiSolv-Exp & 0.66          & 0.92           & 0.12                   \\ \hline 
\end{tabular}
\end{table}

\begin{figure}[h!]
 \begin{subfigure}{0.475\textwidth}
     \includegraphics[width=\textwidth]{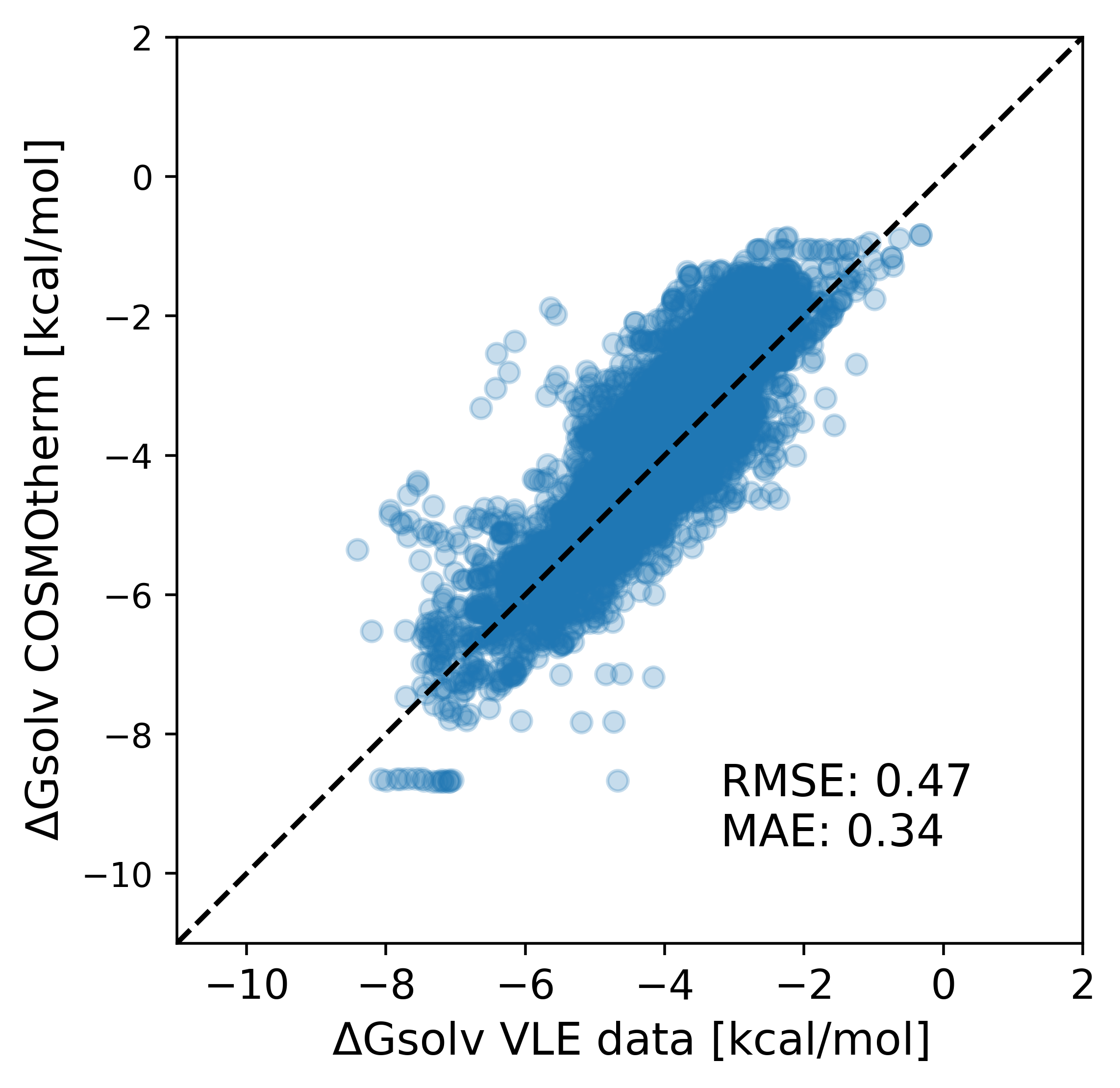}
     \caption{COSMOtherm - Non-aqueous}
     \label{fig:a}
 \end{subfigure}
 \hfill
 \begin{subfigure}{0.475\textwidth}
     \includegraphics[width=\textwidth]{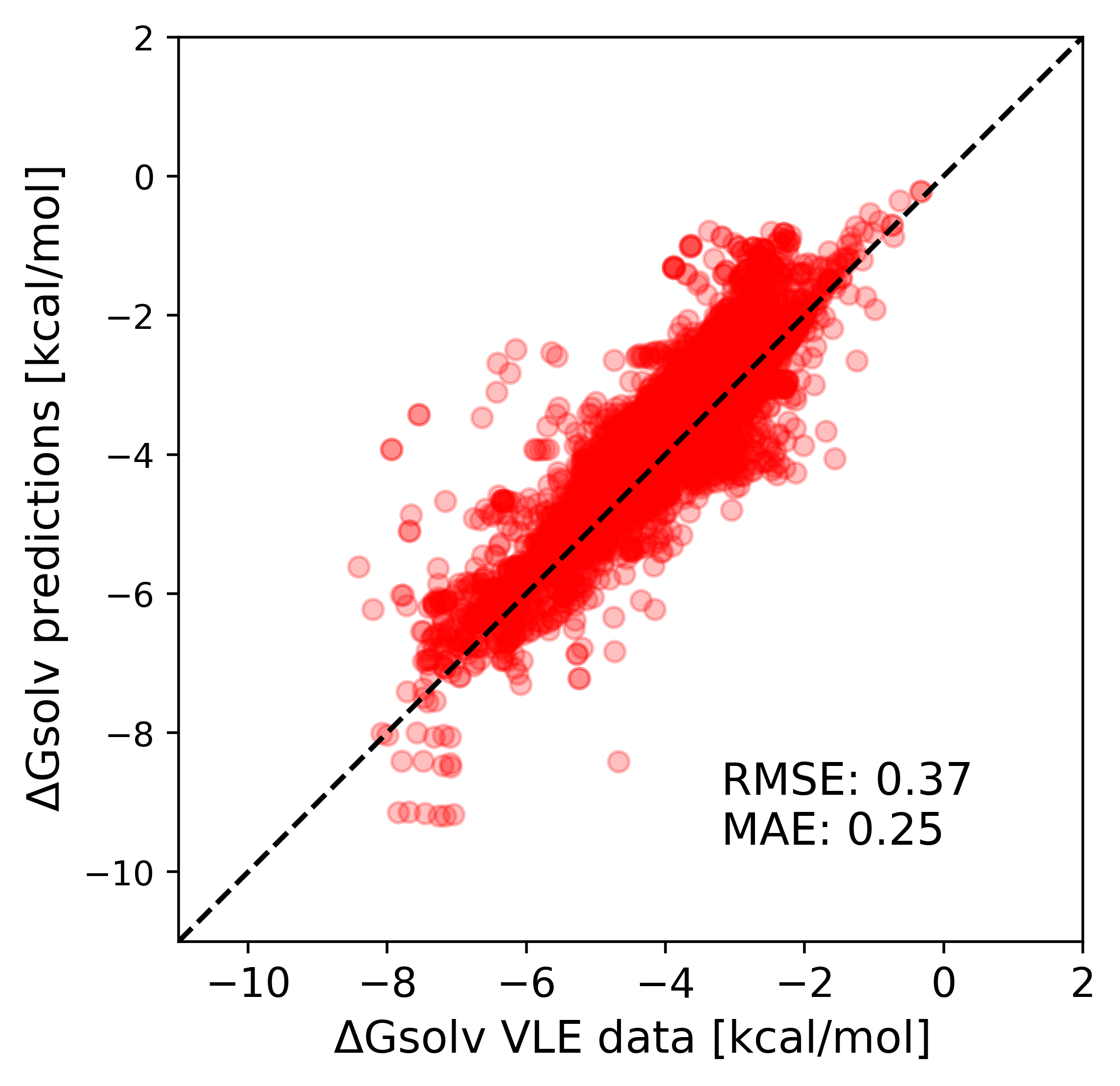}
     \caption{SolProp-mix - Non-aqueous}
     \label{fig:b}
 \end{subfigure}
 
 \medskip
 \begin{subfigure}{0.475\textwidth}
     \includegraphics[width=\textwidth]{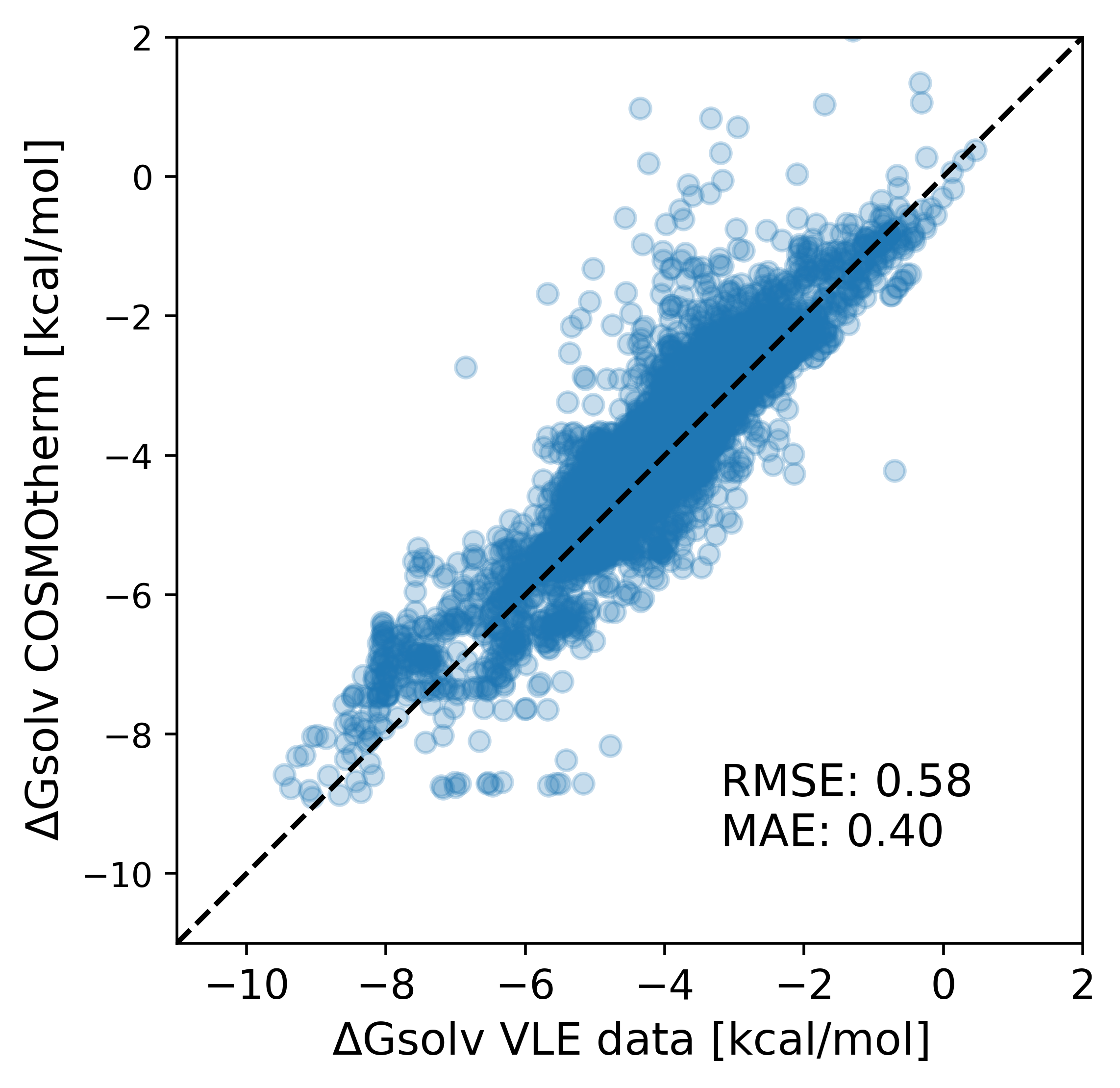}
     \caption{COSMOtherm - Aqueous}
     \label{fig:c}
 \end{subfigure}
 \hfill
 \begin{subfigure}{0.475\textwidth}
     \includegraphics[width=\textwidth]{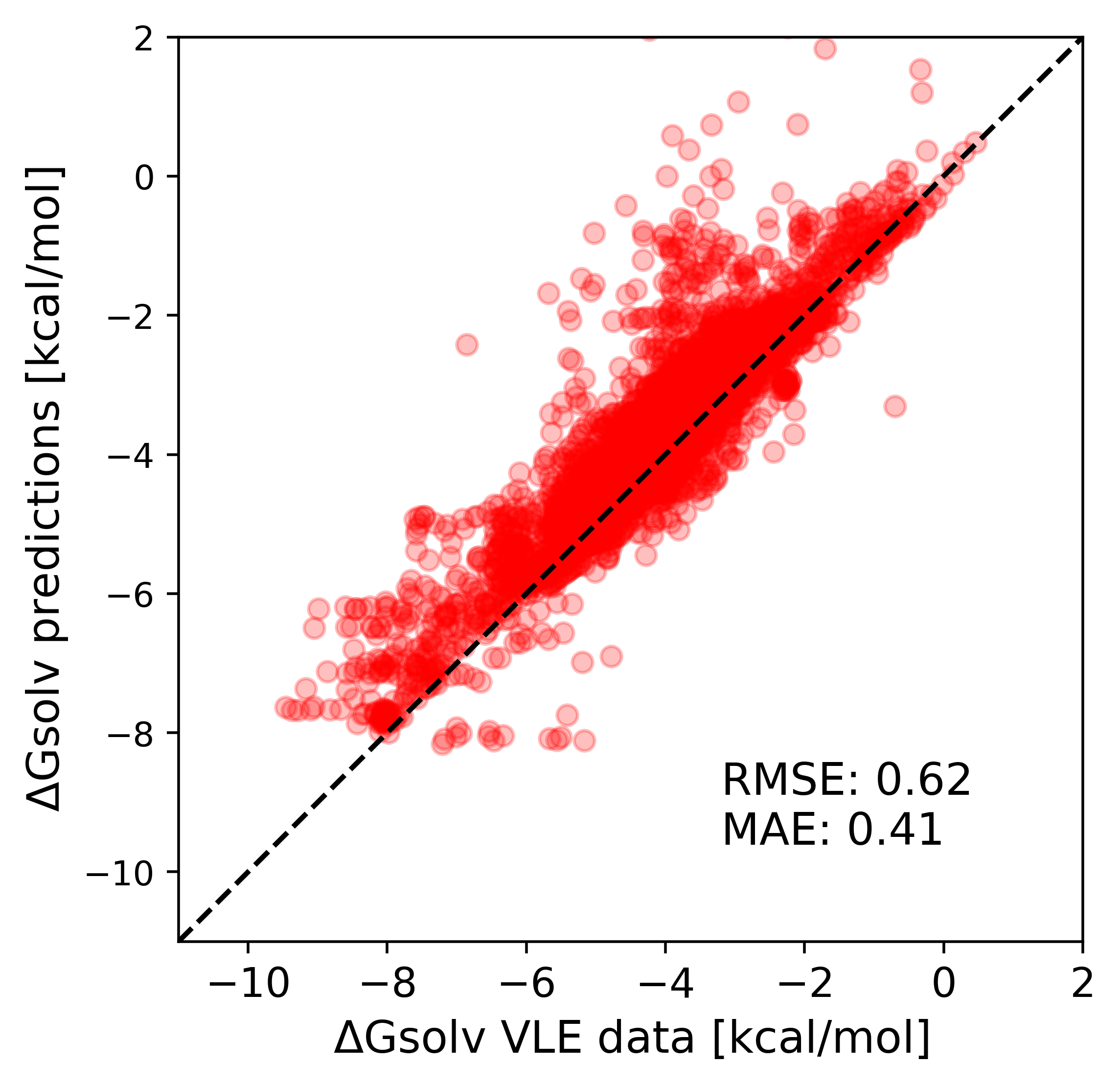}
     \caption{SolProp-mix - Aqueous}
     \label{fig:d}
 \end{subfigure}

 \caption{Four parity plots for the predictions on BinarySolv-Exp dataset, split by whether water is one of the components in the solvent mixture (top) or not (bottom).  COSMOtherm calculations are on the left and fine-tuned SolProp-mix predictions are on the right.}
 \label{fig:splits}
\end{figure}

\begin{table}[h!]
\caption{The performance of different architectures and benchmarks on the roughly 22500 data points in the BinarySolv-Exp database that do not include water as any of the solvents. "QM" in the training data column refers to the merged CombiSolv-QM and BinarySolv-QM database. $\sigma$ refers to the average standard deviation over the test data for the model ensemble.} 
\label{tab:performance_no_water}
\footnotesize
\begin{tabular}{|l|l|l|l|l|}
\hline
Model       & Training data       & MAE(kcal/mol) & RMSE(kcal/mol) & $\sigma$(kcal/mol)  \\ \hline 
COSMOtherm  & -                   & 0.34          & 0.47           & -                   \\ \hline 
SolProp-mix & QM                  & 0.31          & 0.43           & 0.088               \\ \hline 
SolProp-mix & CombiSolv-Exp       & 0.30          & 0.42           & 0.13                \\ \hline 
SolProp-mix & QM \& CombiSolv-Exp & \textbf{0.25} & \textbf{0.37}  & 0.068               \\ \hline 
Concat      & QM                  & 0.30          & 0.42           & 0.11                \\ \hline 
Concat      & QM \& CombiSolv-Exp & 0.28          & 0.40           & 0.089               \\ \hline 
\end{tabular}
\end{table}
th COSMOtherm and the fine-tuned SolProp-mix models were less accurate on aqueous data as compared to non-aqueous data. Parity plots for the performance of both models on the aqeuous and non-aqueous data are shown in \autoref{fig:splits}. For the aqueous data, both models on average predict a value of the solvation free energy that is higher than the experimental value in BinarySolv-Exp. The COSMOtherm values had a Mean Signed Error (MSE) of $0.13\ kcal/mol$ on the aqueous data and a MSE of $0.11\ kcal/mol$ on the non-aqueous data. The SolProp-mix values showed a more pronounced effect with a MSE of $0.33\ kcal/mol$ on the aqueous data and a MSE of $0.17\ kcal/mol$ on the non-aqeuous data. The MSE of $0.39\ kcal/mol$ on the aqueous data shows clearly in \autoref{fig:d}, where the distribution is not center on the parity line.
Add that the fine-tuning did not center the distribution on the parity line.

\begin{figure}[h!]
  \centering
  \includegraphics[width=0.5\linewidth]{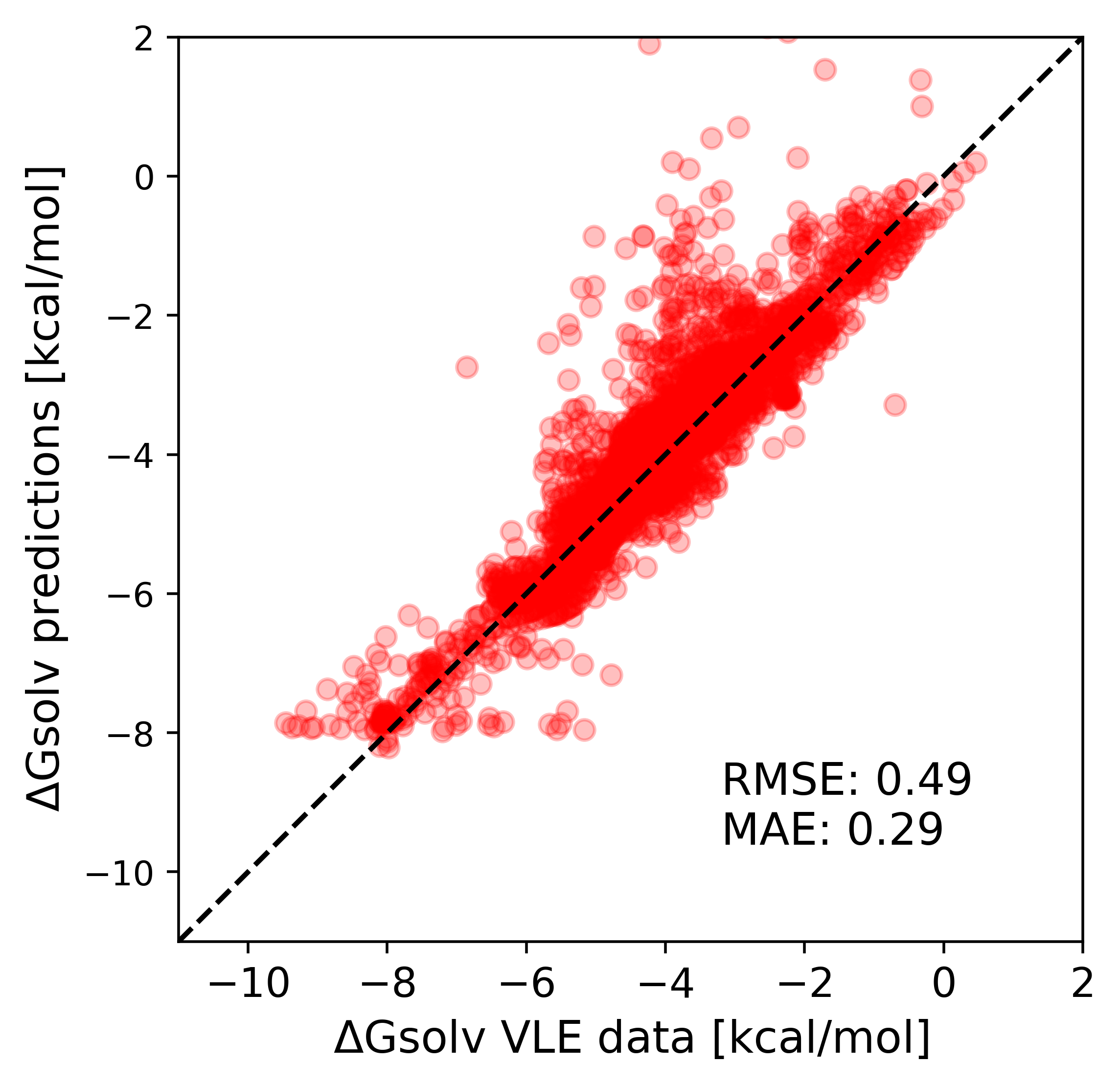}
  \label{fig:sub2}
\caption{A parity plot for the predictions on the BinarySolv-Exp data points, where water is one of the components in the solvent mixture. The SolProp-mix model is here additionally fine-tuned on 300 data points that included water in the mixture.}
\label{fig:water_fix}
\end{figure}

Predictions in aqueous solvents have shown to be more difficult in previous work \cite{VermeireTL,Winter2}. The reason for these inaccuracies is still not fully understood, but could be related to the high polarity of water molecules and thus the stronger interactions with other components in the mixture. Another potential issue is that the QM data only accounts for the thermodynamic solubility of the solutes while during experimental measurements reactions between the aqueous solvent and the solute might increase solubility. This effect was mitigated by filtering out solutes that readily react with water (e.g. acidic and basic species) from the experimental solvent mixture datasets.
We suppose that, as a first step, adding solvent mixtures data that include water to the experimental training set  could improve the accuracy of SolProp-mix. Therefore, we also fine-tuned a QM-data pre-trained SolProp-mix model on CombiSolv-Exp supplemented with 300 water mixture data points from BinarySolv-Exp. These 300 points were points for solutes that only appeared in water mixtures. The predictions for the other water mixture data points in BinarySolv-Exp are shown in \autoref{fig:water_fix}. By inclusion of a small sample of solutes in water mixtures, the SolProp-mix architecture improved the accuracy from an RMSE of $0.62 \ kcal/mol$ to $0.49 \ kcal/mol$ for these data points, which is more comparable to the performance on organic solvent mixtures.

\subsubsection{Solvation prediction for ternary solvent mixtures}
The fine-tuned SolProp-mix model that was trained on data with one or two components in the solvent mixture also allows to make predictions on systems with more components in the solvent mixture. To further explore this possibility, a comparison was made between COSMOtherm calculations and the fine-tuned SolProp-mix model for the experimental data in ternary solvent mixtures (TernarySolv-Exp).

\begin{figure}[h!]
\centering
\begin{subfigure}{.475\textwidth}
  \centering
  \includegraphics[width=\linewidth]{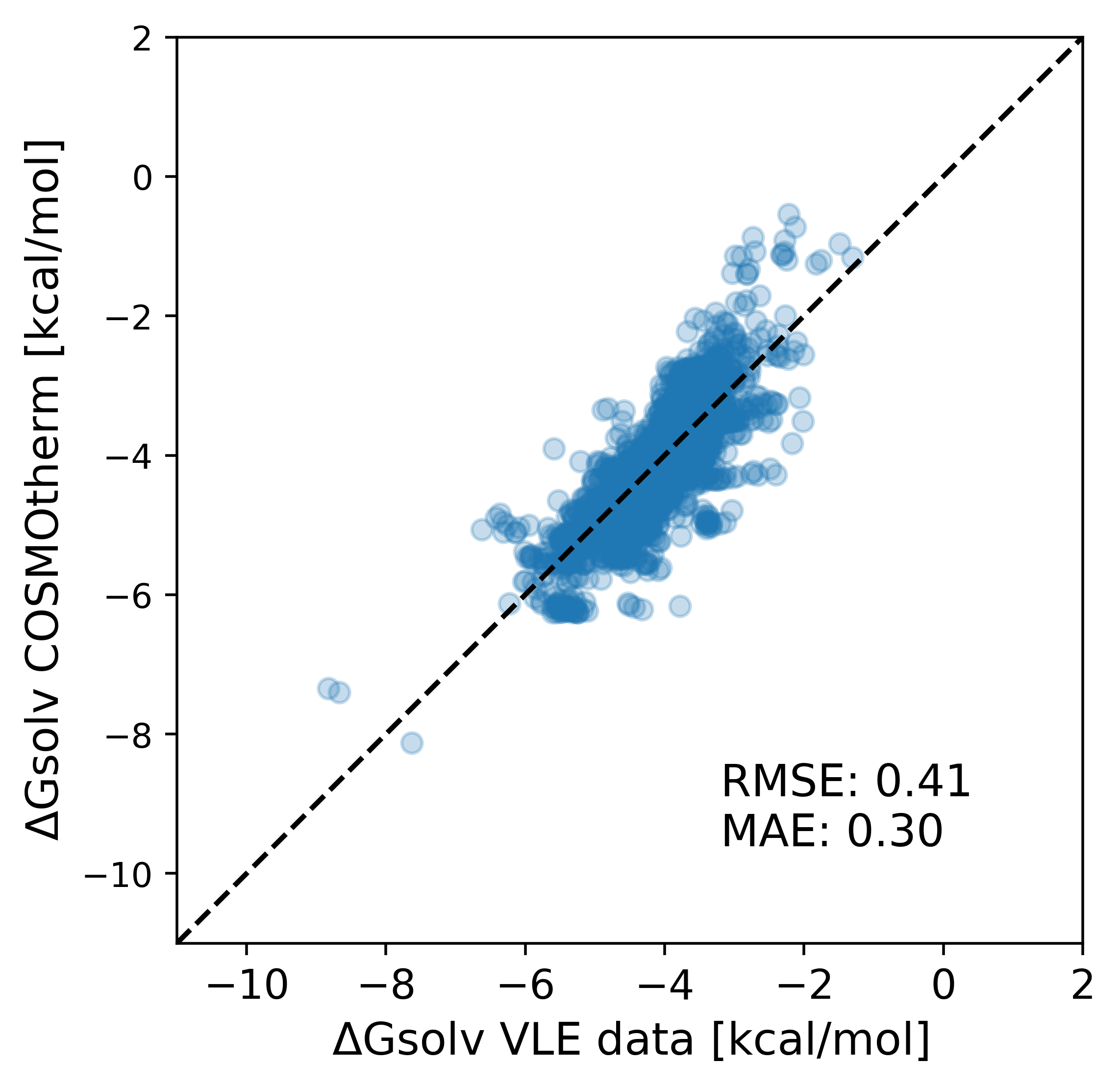}
  \caption{COSMOtherm}
  \label{fig:sub1_water}
\end{subfigure}%
\begin{subfigure}{.475\textwidth}
  \centering
  \includegraphics[width=\linewidth]{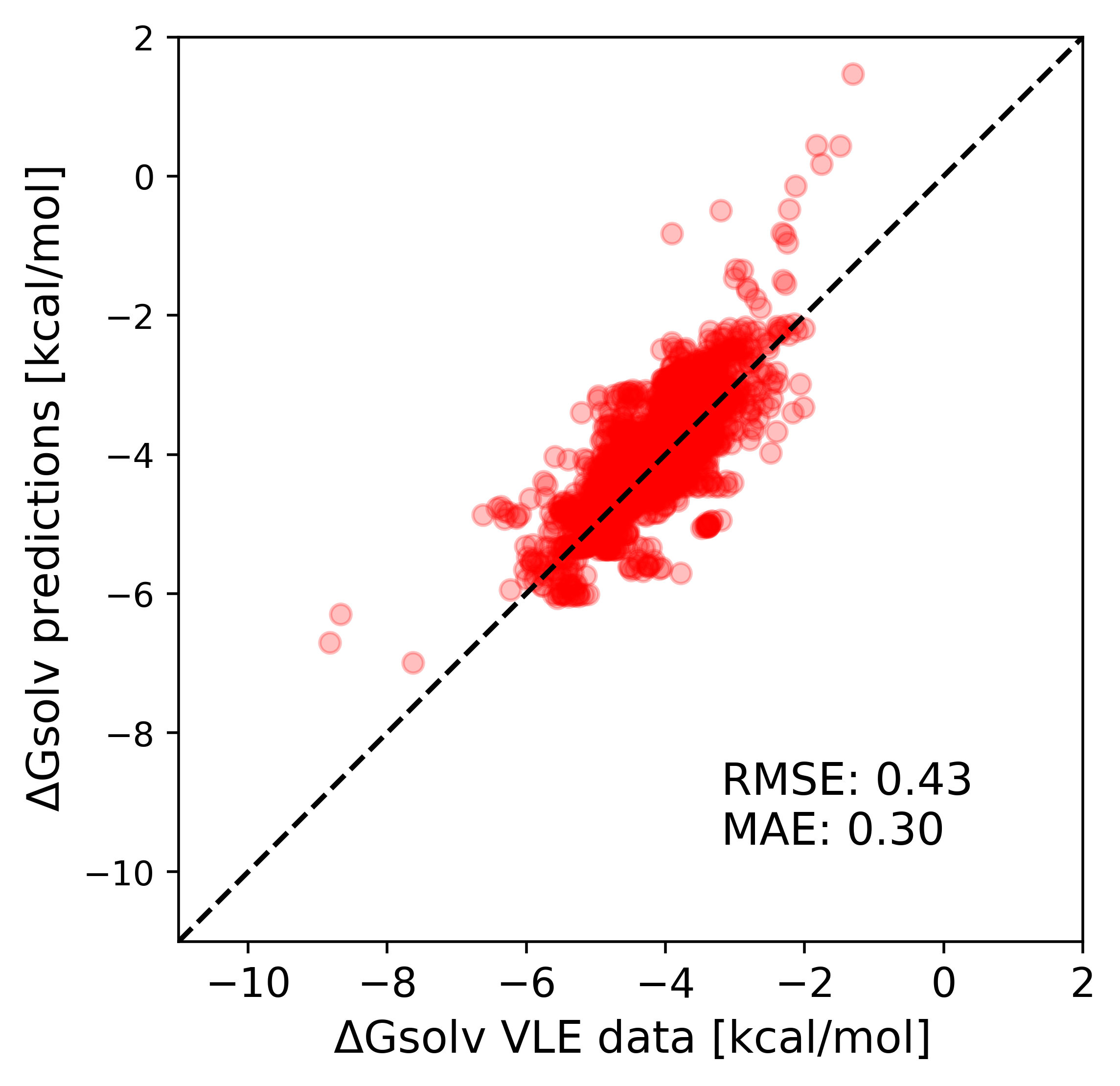}
  \caption{SolProp-mix}
  \label{fig:sub2}
\end{subfigure}
\caption{Two parity plots for the predictions on the TernarySolv-Exp database for COSMOtherm and fine-tuned SolProp-mix implementation.}
\label{fig:ternary_res}
\end{figure}

\autoref{fig:ternary_res} shows two parity plots, one for the calculations made by COSMOtherm and one for the predictions made by the SolProp-mix implementation. A comparison with the parity plots in binary solvents shows that the range of solvation free energies was notably smaller for the ternary solvents, while the model errors were of the same order of magnitude. SolProp-mix and COSMOtherm perform similarly with SolProp-mix having an RMSE of $0.34 \ kcal/mol$ and an MAE of $0.24 \ kcal/mol$ and COSMOtherm having an RMSE of $0.41 \ kcal/mol$ and an MAE of $0.30 \ kcal/mol$. The COSMO-RS theory is known to apply well to multi-component mixtures \cite{klamt}. Therefore, the ability of SolProp-mix to replicate this is an important quality.

\subsubsection{Trend validation for mixture compositions}
\begin{figure}[h!]
\centering
\begin{subfigure}{.49\textwidth}
    \centering
    \includegraphics[width=\linewidth]{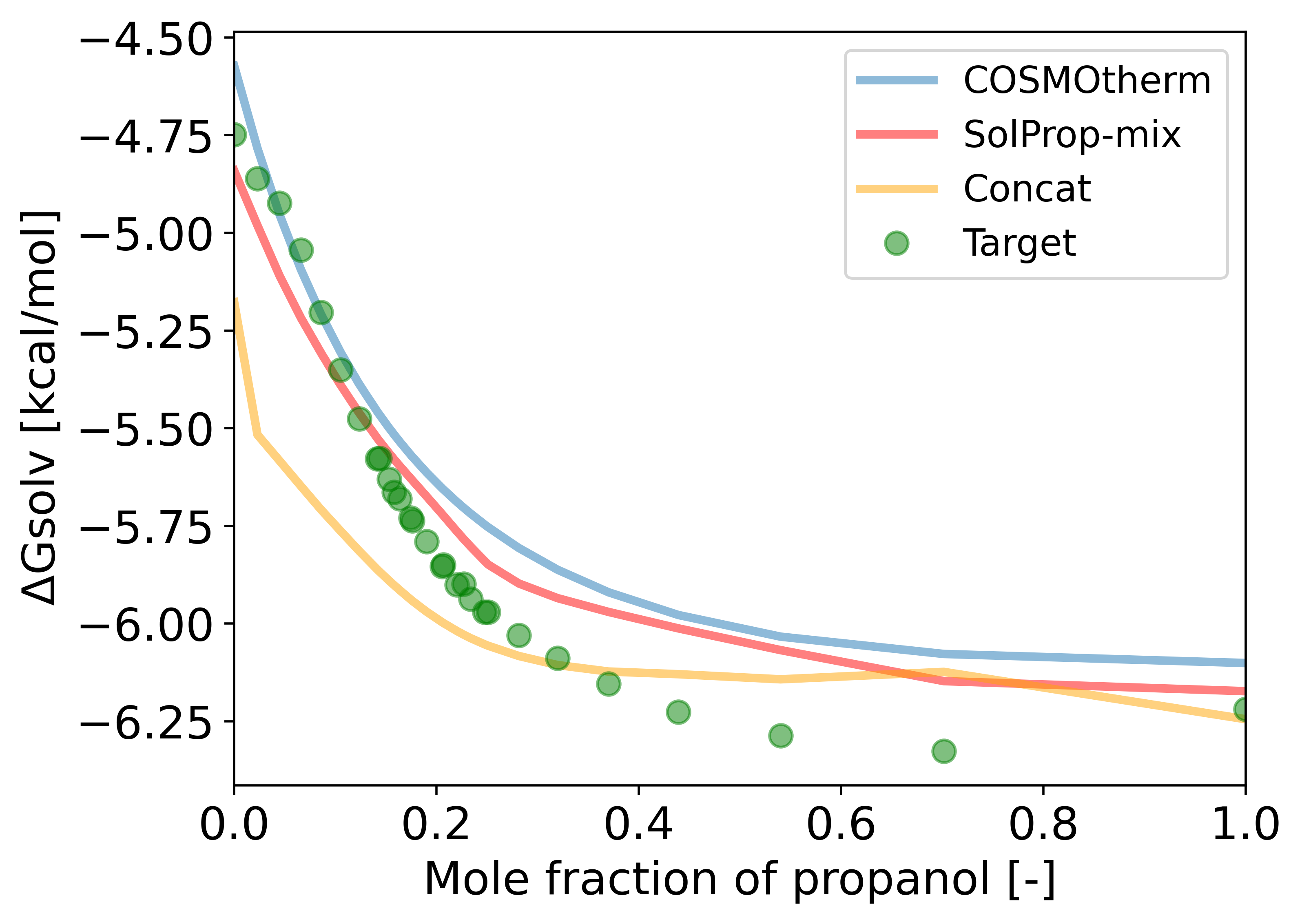}
    \label{fig:1png}
\end{subfigure}%
\begin{subfigure}{.49\textwidth}
    \centering
    \includegraphics[width=\linewidth]{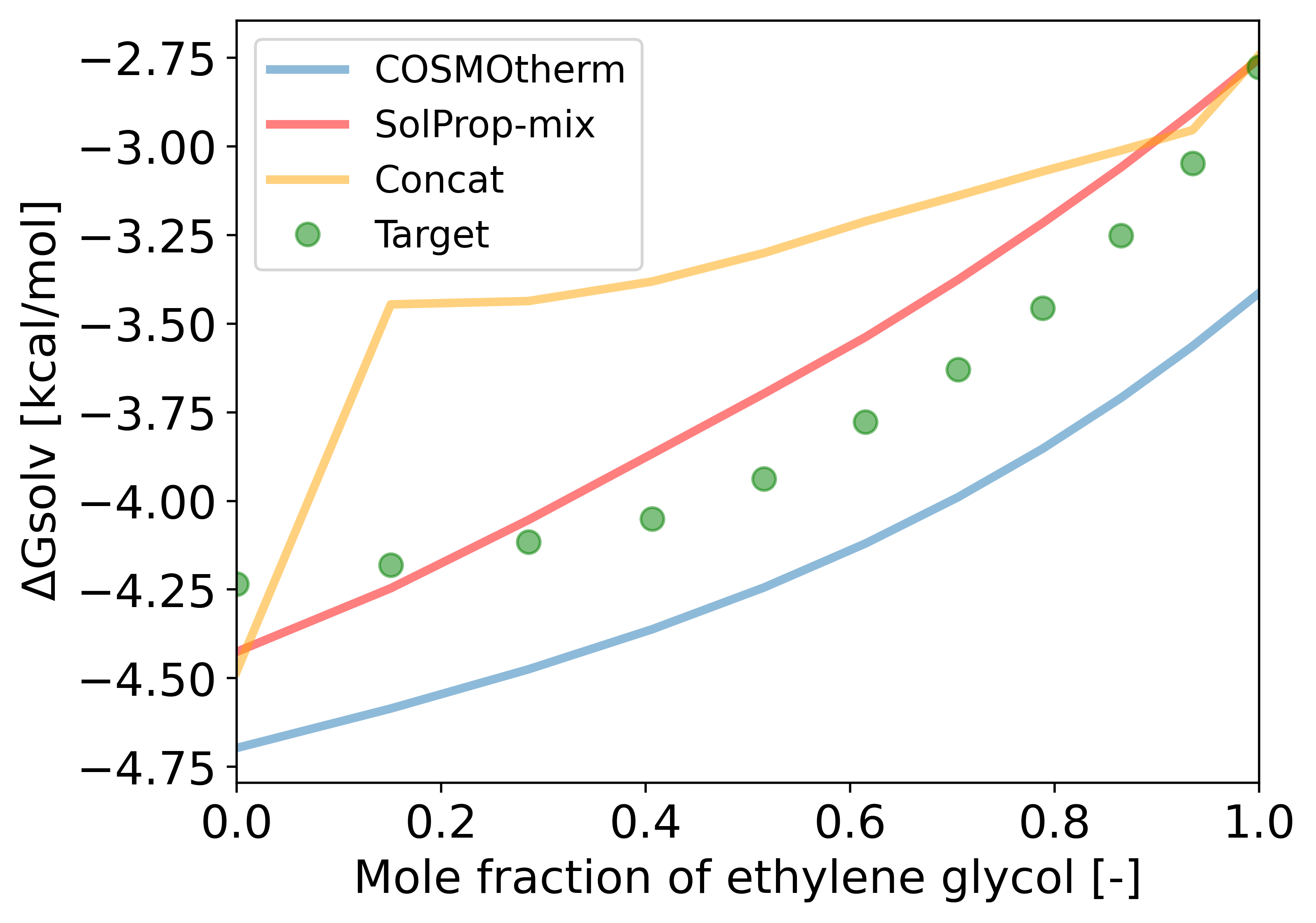}
    \label{fig:3png}
\end{subfigure}
\caption{Solvation free energies in different binary solvent compositions at $298K$ obtained from BinarySolv-Exp. The left panel shows butanol in a mixture of propanol and water, and the right panel shows benzene in a mixture of ethylene glycol and methylpyrrolidone. The blue line shows the calculations made by COSMOtherm and the green points the measurements. The red and orange lines show predictions by the fine-tuned SolProp-mix and Concat models, respectively.}
\label{fig:trend_gsolv}
\end{figure}

When designing new solvents mixtures to optimize reaction rates or separation processes, both the identity and component fraction of each component in the mixture can be optimized. Thus it is important that models for predicting solvation free energy in mixtures capture the behavior of the solvation free energy as a function of the solvent mixture's composition. 
\autoref{fig:trend_gsolv} shows the change of the solvation free energy for two systems of a specific solute in a varying composition of the binary solvent mixture together with the predictions made by the different models of this work. Plots for additional systems are available in the SI. The mixtures in \autoref{fig:trend_gsolv} were selected from the data converted from IDACs as measurements were available across the whole compositional range.

The COSMOtherm calculations capture trends in thermodynamic properties for molecular mixtures accurately, as has been reported previously \cite{cosmo_trends}. As can be seen in \autoref{fig:trend_gsolv}, SolProp-mix captures the composition dependent trends similar to the COSMOtherm calculations. In addition, SolProp-mix showed an improved absolute performance as compared to COSMOtherm, as was expected for non-aqueous solvent mixtures based on the overall performance in \autoref{tab:performance_no_water}.

The neural networks proposed in this work were extensively trained on COSMOtherm calculations, therefore, an adequate implementation should capture these trends similarly. Although this was the case for the SolProp-mix architecture, the Concat architecture is not able to capture these trends accurately. We expect that the non-smooth predictions around $x=0$ and $x=1$ for the Concat architecture comes from an issue with the fine-tuning on monosolvents, showing this architecture is compromised in generalizing to different mole fractions.

\section{Conclusions}

A large QM dataset in binary solvent mixtures (BinarySolv-QM) was created to (1) benchmark the new model architecture against state-of-the-art models and (2) for pre-training the final model to predict solvation free energies in solvents mixtures. 
To test the models developed in this work, we carefully curated experimental datasets of solvation free energies in binary (BinarySolv-Exp) and ternary solvents (TernarySolv-Exp) from data on vapor-liquid equilibria and activity coefficients.

The development of SolProp-mix with the MolPool pooling function represents a significant advancement in modeling solvation free energies in complex solvent mixtures, offering improved accuracy, flexibility, and scalability over existing methods. 

The MolPool pooling function was proposed for the generation of a mixture embedding using embeddings of individual components and their mole fractions. MolPool distinguishes itself from literature ML models for mixtures by its permutational invariance towards the mixture composition and by its ability to process an arbitrary number of components in the mixture. Combined with the SolProp package, this new SolProp-mix software has an improved performance for predicting solvation free energies. It enhances generalization, requires less data, and reduces the required model weights as compared to state-of-the-art architectures (\textit{i.e.} concatenation and solvGNN). 

The MolPool pooling function creates the possibility to train a model on mixtures with an arbitrary number of components. In addition, it also has the ability to extrapolate to mixtures with more components than were present in the training data, which is especially crucial because often data for mixtures with more components are less available. We observed an improved performance in predicting solvation free energy for binary and ternary solvents by fine-tuning our QM-based model on experimental data in monosolvents. This clearly demonstrates the high potential of pooling molecule embeddings. 
On the BinarySolv-Exp data, which is our newly compiled dataset from vapour-liquid equilibrium data and infinite dilution activity coefficients, the RMSE was $0.46 \ kcal/mol$ and the MAE $0.29 \ kcal/mol$. For the TernarySolv-Exp data, the RMSE was $0.40 \ kcal/mol$ and the MAE $0.28 \ kcal/mol$. The errors were comparable to COSMOtherm calculations that were used as a benchmark. In addition, when the comparison was limited to non-aqueous systems, enhanced performance of SolProp-mix as compared to COSMOtherm was observed.

In general, it was observed that the performance of both our models and COSMOtherm calculations for mixtures that include only organic solvents is better as compared to aqueous solutions. Similar observations were made in other works and it remains unclear what the exact reason is for these higher discrepancies in aqueous solutions.
Besides achieving a good absolute performance, SolProp-mix is accurate at predicting trends of solvation free energies in varying solvent mixtures, which is critical, for example, for formulating promising solvent mixtures in a design phase.
To conclude, the MolPool pooling function can be applied to predict properties of molecular mixtures, extending the application range of graph neural networks to a whole variety of new thermochemical systems.

\section*{Acknowledgments}
The authors would like to thank Yunsie Chung for her help with the generation of the quantum chemical database for the enthalpy of solvation in mixtures. The authors acknowledge the Fonds Wetenschappelijk Onderzoek (FWO) for funding (G021924N). Resources and services used in this work were provided by the VSC (Flemish Supercomputer Center), funded by the Research Foundation - Flanders (FWO) and the Flemish Government. E. A. acknowledges
the support of the Ibn Rushd Postdoctoral Fellowship Program, administered by the King
Abdullah University of Science and Technology (KAUST).
The authors acknowledge the Machine Learning for Pharmaceutical Discovery and Synthesis Consortium (MLPDS), and the DARPA Accelerated Molecular Discovery (AMD) program (DARPA HR00111920025) for funding. 
This research used HPC resources to perform quantum calculation of the National Energy Research Scientific Computing Center (NERSC), a U.S. Department of Energy Office of Science User Facility operated under Contract No. DE-AC02-05CH11231.

\section{Supplementary data}
\label{sec:sample:appendix}
The SolProp package can be found at the following link: \newline\url{https://gitlab.kuleuven.be/creas/vermeiregroup/solprop}. Datasets, model weights, and a static version of the code can also be found on Zenodo: \newline \url{doi:10.5281/zenodo.14238055}.

\bibliographystyle{elsarticle-num}
\bibliography{achemso-demo}

\end{document}


\appendix
\section{The creation of the quantum chemistry database using active learning}
For the training of the neural networks, a large and diverse dataset of solvation free energies in solvent mixtures based on quantum chemistry methods was created. Given the same solute database as in our previous work \cite{VermeireTL} (11,000 solutes), along with 931 combinations of miscible solvents \cite{SigmaAldrich} and 9 mole fractions of each solvent, there are 92 million potential combinations available for calculation. 
This would be an infeasible number of data points to calculate and to train the graph neural networks on. While having a large database is valuable, the diversity of the data is even more crucial \cite{Heid}. To address this, we employed an active learning approach—a machine learning technique designed to optimize model performance with a minimal amount of labeled data.

\begin{figure}[h!]
    \centering
    \includegraphics[scale=0.75]{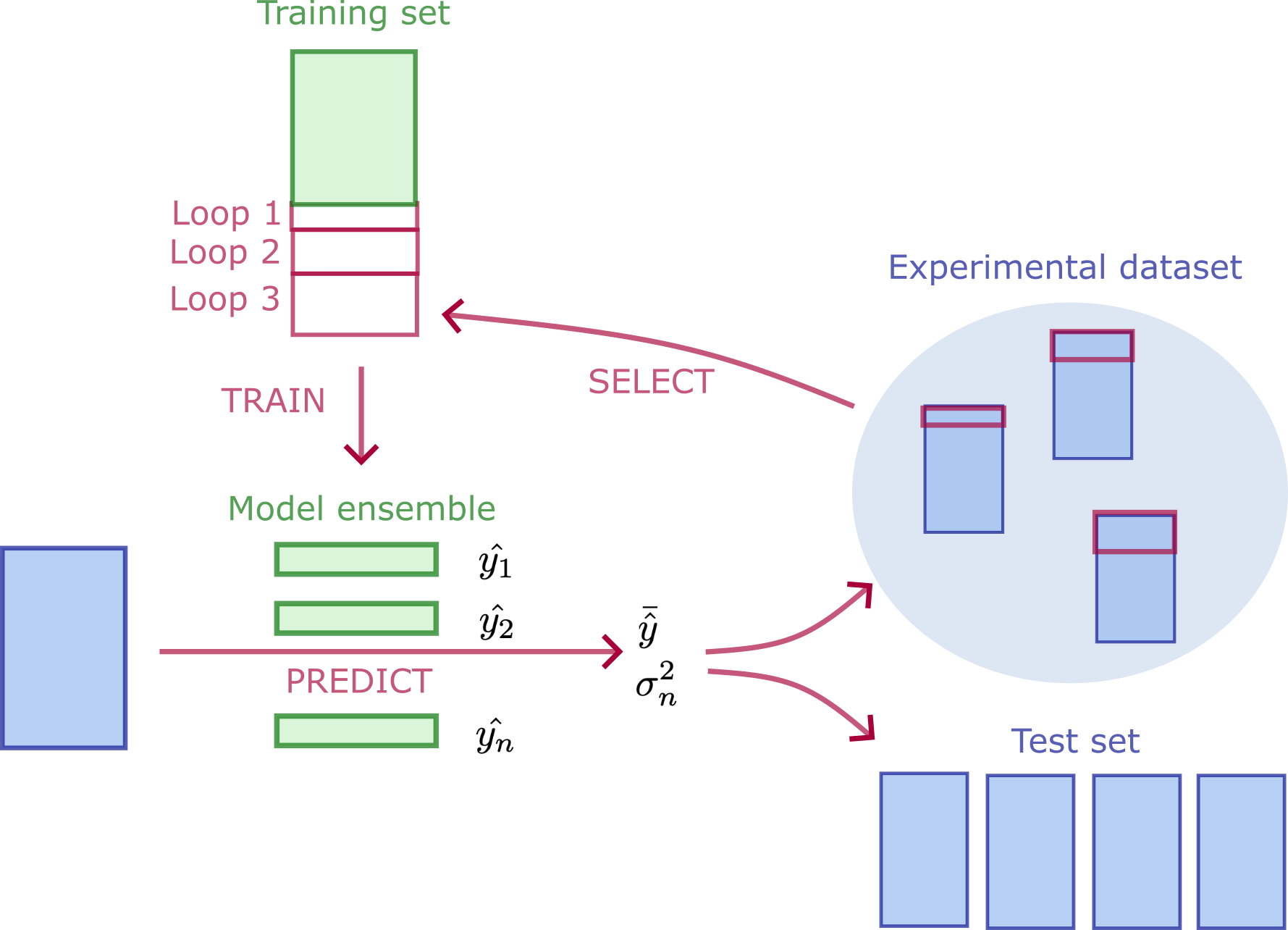}
    \caption{An overview of the active learning loop is shown.}
    \label{fig:active_learning}
\end{figure}

In this process, the active learning algorithm identified specific data points, which were then calculated using COSMOtherm. During each round of active learning, new entries were selected from a list of possible data points (ref. the experimental dataset) for which COSMO-RS \cite{cosmors} calculations were automatically carried out using the COSMOtherm software \citep{Dassault}. Each iteration of the active learning loop (see \autoref{fig:active_learning}) began by adding these new calculations to the training data. Following this, a model ensemble was trained on the updated dataset, and the acquisition function identified additional data points for further analysis.

In the design of the acquisition function, different ways of exploration were employed. Our aim was to develop a widely applicable model to predict $\Delta G_{solv}$ in solvent mixtures, therefore the acquisition function was designed based on uncertainty measures and the uniqueness of molecules included. 
The number of calculations to be performed in every active learning loop was determined as 10\% of the size of the training set. The subset of experimental data used for predictions and selecting new entries was deliberately limited in size compared to the full experimental design space to minimize the computational cost of the prediction step.
The subset of the experimental data had 100 times more entries as compared to the number of data points that will be added and is re-constructed every iteration by a random sampling from all remaining combinations of solutes, solvents, and mole fractions in the experimental design space.   

To evaluate how data was selected based on functions that relate to the epistemic uncertainty, the performance of three different acquisition functions was compared. The acquisition functions were designed based on: the ensemble variance ($s^2$), the scaled ensemble variance ($s^2_{scaled}$), and a combination of the ensemble variance and solute uniqueness ($s^2+\exists!m\in\mathbb{M}_{exp}$). The ensemble variance (Eq. \ref{EQU:ens_var}), or the variance contribution to the total epistemic uncertainty \cite{Heid}, is the model uncertainty approximated by the variance between the predictions from an ensemble of 30 models. With $\hat{y_{i}}$ the prediction of model $i$ in the ensemble of $N_{ens}$ models for input $X_n$, and $\overline{\hat{y}}$ the average prediction of the $N_{ens}$ models.

\begin{equation}
    s^2(X_n) = \frac{\sum_{i}^{N_\mathrm{ens}} (\hat{y_{i}}(X_n) - \overline{\hat{y}}(X_n))^2}{N_\mathrm{ens}}
    \label{EQU:ens_var}
\end{equation}

The scaled version of the ensemble variance is given by Eq. \ref{EQU:ens_var_scaled}. Scaling was applied since preliminary results have shown that the ensemble variance is larger for larger absolute values of the solvation free energy. 
To improve data acquisition for solute/solvent combinations with lower solvation free energies, the ensemble variance was scaled with the sum of the absolute value of the average prediction of the model ensemble and the average prediction of all data in the experimental dataset (see Eq. \ref{EQU:ens_var_scaled}). The latter term was added, since scaling based on the average prediction resulted in too high uncertainties when the predicted solvation free energy is too close to zero.

\begin{equation}
    s^2_{scaled} (X_n) = \frac{s^2(X_n)}{\sqrt{\overline{\hat{y}}(X_n)^2}+\frac{1}{N_\mathrm{data}}\sum_{n}^{N_\mathrm{data}} \sqrt{\overline{\hat{y}}(X_n)^2}} 
    \label{EQU:ens_var_scaled}
\end{equation}

When the ensemble variance and the solute uniqueness were combined, the new data points were selected based on the highest ensemble variance with the constraint that there were no duplicates within the added solutes. 
With this approach, we attempted to reduce the bias component of the epistemic uncertainty in addition to the variance. 
Separately, a random acquisition function was used as a reference to evaluate the overall performance of the active learning algorithm.
The databases stemming from the procedures using the different acquisition functions were finally merged to form the database BinarySolv-QM.

\section{The curation of the experimental databases}
\subsection{Theoretical foundation for the data curation}
To compare the performance of our final models to the COSMOtherm implementation of COSMO-RS, a set of experimental solvation free energy ($\Delta G_{solv}$) data in binary and ternary solvents was curated. $\Delta G_{solv}$ was not measured directly but calculated from experimental three- and four-component vapor liquid equilibrium (VLE) data and binary solvent infinite dilution activity coefficient (IDAC) data. 

As a starting point for calculating the experimental $\Delta G_{solv}$ in a mixed solvent, we used \autoref{eq:g_solv_fugacity}, which is derived by Moine et al. \cite{Moine} from the Ben-Naim definition of solvation \cite{BenNaim}. This equation requires the vapor and liquid phase molar densities, $\rho_{vap}$ and $\rho_{liq}$, and solute's liquid phase fugacity coefficient ($\varphi_{i,\,liq}$). These are functions of temperature $T$, pressure $P$, and vapor or liquid molar compositions $\mathbf{y}$ and $\mathbf{x}$. $R$ is the universal gas constant.

\begin{equation} \label{eq:g_solv_fugacity}
    \Delta G_{i,\,solv}(T,P,\mathbf{x}) = RT \ln\left[\varphi_{i,liq}(T,P,\mathbf{x})\frac{\rho_{vap}(T,P,\mathbf{y})}{\rho_{liq}(T,P,\mathbf{x})}\right]
\end{equation}

\subsubsection{Solvation free energy from vapor liquid equilibrium measurements}
In multi-component VLE measurements, the compositions of a vapor and liquid phase in equilibrium are measured at a specified temperature and pressure. Because the two phases are in equilibrium, each component's fugacity in the vapor phase is equal to its fugacity in the liquid phase. Thus, from the definition of fugacity coefficients \cite{Smith2005}, \autoref{eq:fugacity_coefficients} is true for each VLE measurement. $x_i$ and $y_i$ are the mole fractions of component $i$ in the vapor phase and liquid phase, respectively. 

\begin{equation} \label{eq:fugacity_coefficients}
 P\cdot y_i \cdot \varphi_{i, gas} = P\cdot x_i \cdot \varphi_{i, liq}
\end{equation}

This equation is readily rearranged to give an expression for the liquid phase fugacity coefficient of component $i$. Further, if the pressure of the VLE measurement is low enough, the vapor phase can be assumed an ideal gas. Then, the vapor phase fugacity coefficient of component $i$ is unity and \autoref{eq:fugacity_coefficients} simplifies to \autoref{eq:fug_coef_x_y}. Moreover, the assumption of an ideal vapor phase means the ideal gas equation of state can be used to calculate the vapor phase molar density. 

\begin{equation} \label{eq:fug_coef_x_y}
    \varphi_{i,liq} = \frac{y_i}{x_i}
\end{equation}

Combining \autoref{eq:g_solv_fugacity} and \autoref{eq:fug_coef_x_y} and using the ideal gas equation of state to calculate the vapor phase molar density gives \autoref{eq:VLE_gsolv}. Note that the numerator is equal to the molar concentration of component $i$ in the vapor phase and the denominator is equal to the molar concentration of the component in the liquid phase. This shows that \autoref{eq:VLE_gsolv} is similar to the expression by Ben-Naim \cite{BenNaim}, which relates the solvation free energy to the equilibrium ratio of the molecular densities of a solute in gas and liquid phases, except that here a molar scale is used.

\begin{equation} \label{eq:VLE_gsolv}
 \Delta G_{i,\,solv}(T,P,\mathbf{x}) = RT \ln\left[\frac{y_i\,P/RT}{x_i\,\rho_{liq}(T,P,\mathbf{x})}\right]
\end{equation}

\autoref{eq:VLE_gsolv} can be used to calculate the solvation free energy for each component in the liquid phase that is present at the VLE measurement's temperature and pressure.
In this work however, we limit our scope to infinite dilution solvation free energies, $\Delta G^\infty_{i,\,solv}$. Thus, we make the additional assumption that interactions between solute molecules in the liquid phase do not affect the solute's solvation energy. This assumption holds best if either the interaction between two solutes is similar to the interaction between a solute and the solvent, or the solute is sufficiently dilute. With this assumption, our final equation for converting VLE measurements to solvation free energy data is \autoref{eq:VLE_gsolv_final}

\begin{equation} \label{eq:VLE_gsolv_final}
 \Delta G^\infty_{i,\,solv}(T,P,\mathbf{x}) =  RT \ln\left[\frac{y_i}{x_i}\frac{P/RT}{\rho_{liq}(T,P,\mathbf{x})}\right] 
\end{equation}

\subsubsection{Solvation free energy from activity coefficients}
Activity coefficient measurements indicate the non-ideal behavior of a component in the liquid phase. The fugacity of component $i$ in the liquid phase is related to its activity coefficient, $\gamma_{i,liq}$, by \autoref{eq:activity_fugacity} where $f_{pure-i, liq}(T,P)$ is the fugacity of component $i$ in a pure liquid phase of only component $i$ at the measurement's temperature and pressure. 

\begin{equation} \label{eq:activity_fugacity}
    f_{i,liq}(T, P, x) = f_{pure-i, liq}(T,P) \cdot x_i \cdot \gamma_{i,liq}(T,P,x)
\end{equation}

$f_{pure-i, liq}(T,P)$ is related to the vapor pressure of component $i$, $P_i^{sat}$ by \autoref{eq:pure-liquid-fugacity}. This equation is exact but relies on the solute's saturation fugacity coefficient, $\varphi_i^{sat}(T)$, and the Poynting factor. The saturation fugacity coefficient only deviates from unity at very high temperatures and the Poynting factor only deviates significantly from unity at very high pressures. Thus, under the conditions in the scope of this work, the pure liquid fugacity of the solute is well approximated by its vapor pressure. 

\begin{equation} \label{eq:pure-liquid-fugacity}
  \begin{aligned}
    f_{pure-i, liq}(T,P) = P_i^{sat}(T)\cdot \varphi_i^{sat}(T)\cdot \exp\left[\frac{1}{RT}\int_{P_i^{sat}(T)}^P\frac{dP}{\rho_{pure-i,liq}(T,P)}\right] \\
    \approx P_i^{sat}(T)
  \end{aligned}
\end{equation}

By combining \autoref{eq:activity_fugacity} with \autoref{eq:pure-liquid-fugacity} and applying the definition of fugacity coefficient, \autoref{eq:fug_coef_gamma} is derived which relates the liquid phase fugacity coefficient of component $i$ to its activity coefficient. This can then be combined with \autoref{eq:g_solv_fugacity} to get \autoref{eq:act_gsolv} which relates the activity coefficient to the solvation free energy.

\begin{equation} \label{eq:fug_coef_gamma}
    \varphi_{i,liq} = \gamma_{i, liq}(T,P,x)\frac{P_i^{sat}(T)}{P}
\end{equation}

\begin{equation} \label{eq:act_gsolv}
    \Delta G_{i,\,solv}(T,P,\mathbf{x}) = RT \ln\left[\gamma_{i, liq}(T,P,\mathbf{x})\frac{P_i^{sat}(T)/RT}{\rho(T,P,\mathbf{x})}\right]
\end{equation}

\autoref{eq:act_gsolv} applies to activity coefficients of species at any concentrations. In this work however, only infinite dilution solvation free energies, $\Delta G^\infty_{i,\,solv}$, were within scope. We thus limited our data curation to infinite dilution activity coefficients, $\gamma^\infty$, and used \autoref{eq:act_gsolv_final} instead.

\begin{equation} \label{eq:act_gsolv_final}
    \begin{aligned}
    \Delta G^\infty_{i,\,solv}(T,P,\mathbf{x}) 
    = RT \ln\left[\gamma^\infty_i(T, \mathbf{x})\frac{P_i^{sat}(T)/RT}{\rho_{liq}(T,P,\mathbf{x})}\right]
    \end{aligned}
\end{equation}

\subsection{Experimental data selection}
Data were selected from the Dortmund Data Bank, which is the most comprehensive collection of thermodynamic data reported in the literature.
Three- and four-component VLE data as well as IDACs in binary solvent were extracted from the Dortmund Data Bank (DDB) \cite{Dortmund} with an academic license. Only components labeled as "Normal" by the DDB were selected. The components labeled "Salts", "Adsorbents", and "Polymers" were excluded. The conversion of VLE and IDAC data to $\Delta G^\infty_{solv}$ requires solvent liquid densities and solute vapor pressures. Correlations from the DIPPR 801 database \cite{dippr} were used to calculate these properties. These correlations have the benefit of being evaluated by several academic and industrial experts \cite{how_to_use_dippr}.

\subsubsection{The VLE data curation process}
The general process to convert experimental VLE data to $\Delta G^\infty_{solv}$ was the same for both the three-component and four-component VLE databases and included the following steps:
\begin{enumerate}
    \item Extract data from the DDB.
    \item Remove data points that are missing any of temperature, pressure, vapor phase compositions, and liquid phase compositions.
    \item Remove data points that include metals, ionic liquids, and non-metal salts.
    \item Remove data points where components may react with each other.
    \item Remove data points measured at temperatures above a max temperature cutoff. 
    \item Convert component DDB names to InChIs and remove any data points that fail this conversion.
    \item Convert VLE data points to solute-solvent data points. For each VLE data point, check each component and if it is not water and has a liquid phase mole fraction below a dilute cutoff, consider it a solute and the other components as co-solvents. A single VLE data point may produce zero, one, or multiple solute-solvent data points.
    \item Calculate the solvent density for each solute-solvent data point and remove any data points where the solvent density cannot be calculated.
    \item Calculate $\Delta G^\infty_{i,\,solv}$.
    \item Remove a limited number of suspicious points a posteriori.
    \item Group data points into separate datasets for monosolvent, binary solvent, and ternary solvent.
\end{enumerate}

\begin{figure}[]
    \centering
    \includegraphics[width=0.95\textwidth]{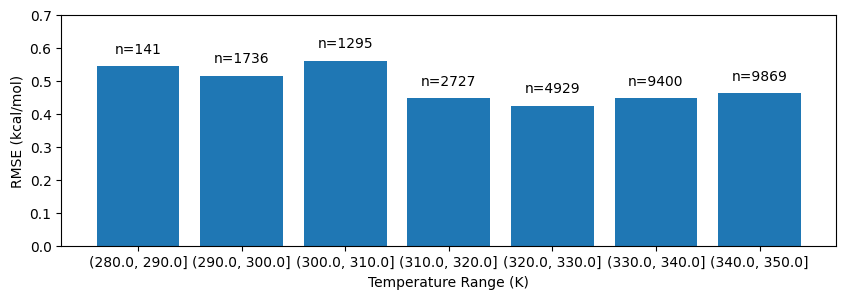}
    \caption{Root mean squared error of SolProp-mix's predictions on BinarySolv-Exp. Data points are binned by the temperature values. Above each bar is the number of data points in that bin.}
    \label{fig:error_vs_binnedT}
\end{figure}

\begin{figure}[]
    \centering
    \includegraphics[width=0.95\textwidth]{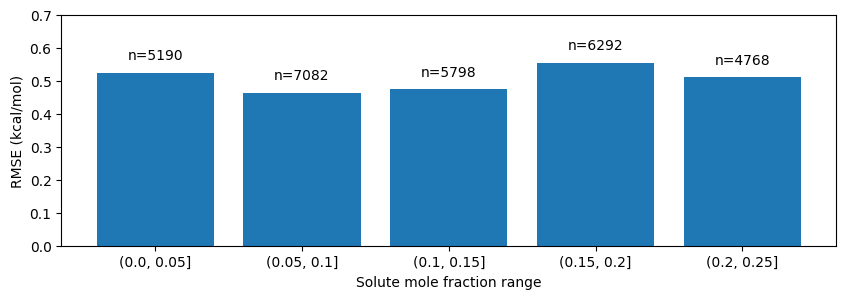}
    \caption{Root mean squared error between SolProp-mix's predictions and infinite dilution solvation free energies calculated using COSMOtherm. The solute-solvent-temperature combinations are taken from the data points in BinarySolv-Exp from VLE data. Data points are binned by the experimental liquid mole fraction of the solute. Above each bar is the number of data points in that bin.}
    \label{fig:error_vs_binnedx}
\end{figure}

Applying this general process in this work also included the following details: 

\begin{enumerate}[Step 1:]

    \item Not all data points in the DDB are licensed for academic use. Our license was an academic one.

    \item There are a variety of VLE experiments that preclude or do not require measuring one of temperature, pressure, vapor composition, or liquid composition. Examples of these include trace component enrichment \cite{trace1997} (no temperature), head-space gas chromatography \cite{doi:10.1021/je00039a031} (no pressure), bubble point measurements \cite{Valtz_Laugier_Richon_1987} (no vapor composition), and dew point measurements \cite{1971122} (no liquid composition). 

    \item The training data for the machine learning models in this work did not include any metals, ionic liquids, or ions. These species are out of the scope for a test set to evaluate the models. 

    \item If any components in a VLE system also form a reactive equilibrium, this would confound the effect of solvation energies. This work considered systems with both water and any of a base, an acid, or formaldehyde as reacting. 

    \item The machine learning models in this work predict both solvation free energy and solvation enthalpy at $298 \ K$. A temperature correction is applied to the solvation free energy at $298 \ K$ by using the solvation enthalpy at $298 \ K$, which assumes the solvation enthalpy does not change with temperature. This assumption breaks down at higher temperatures. A max temperature cutoff of 350 K was selected in this work in accordance with a previous work \cite{VermeireSolu} which assumed that dissolution enthalpy was temperature invariant up to 350 K. It was expected that at higher temperatures SolProp-mix's error would increase as the assumption of constant solvation enthalpy became less valid. Instead the error did not have a clear relationship with with temperature, as shown in \autoref{fig:error_vs_binnedT}. 

    \item InChIs were chosen to identify each species because they are unique for each molecule and standardized. The DDB identifies each species by a common name and DDB specific number. The DDB also reports the CAS Registry Number for many of the species. To convert the DDB species identifiers to InChIs the following steps were applied:
\begin{enumerate}
    \item Remove/replace unidentified characters from species names (e.g. brackets of this form $<>$ are removed and $\alpha$ is converted to alpha).
    \item Map the species name to InChI using PubChem.
    \item Map the CAS number, if available, to InChI using PubChem.
    \item Map the species name to InChI using NIH cactus chemical identifier.
    \item Map the CAS number, if available, to InChI using NIH cactus chemical identifier.
    \item Take the majority vote of the four converted InChI strings.
    \item Manually inspect failed cases.
\end{enumerate}

    \item Calculating solvation free energy from VLE data requires the solute to be in both the vapor and liquid phases at finite concentrations. $\Delta G^\infty_{solv}$ is approximated by assuming the solute's concentration is low enough that the solvation energy calculated is equal to the infinite dilution solvation free energy. We assume this assumption holds for all points where the solute has a liquid mole fraction lower than a cutoff value. We do not use a method based on Raoult's law as has been done elsewhere \cite{Moine} because this method does not extend simply to the case of multiple co-solvents. 

To determine a value for this cutoff, the binary solvent solvation free energies were calculated for all species that had a liquid phase mole fraction less than an arbitrary maximum value of 0.25. Then COSMOtherm was used to calculate the infinite dilution solvation free energy for each of the solute-solvent data points, with the mole fraction of the solute set to zero and the mole fractions of the solvents scaled to sum up to one as show in \autoref{eq:scale_mol_fraction}. It was expected that disagreement between the experimental solvation energy and the COSMO-RS infinite dilution solvation energy would increase as the solute's liquid phase mol fraction increased. Instead the agreement between the two energy values did not have a clear relationship with solute concentration, as shown in \autoref{fig:error_vs_binnedx}. In accordance with this result, 0.25 was used as the dilute solute concentration cutoff in this work. Chung et al. similarly found no obvious correlation between solute mole fraction and the errors of their thermodynamic model \cite{chung_temperature-dependent_2020}.

\begin{equation} \label{eq:scale_mol_fraction}
    x_{solvent_j,scaled}' = \frac{x_{solvent_j}}{\sum\limits_{j} x_{solvent}} = \frac{x_{solvent,j}}{1 - x_{solute}}
\end{equation}

    \item While experimental data for the density of mixtures of solvents would be preferred, such experimental data is sparse. Instead we used a weighted average of the individual solvent densities, which are calculated from correlations available in DIPPR. The weighted averaging was done using \autoref{eq:density} which uses the scaled solvent mole fractions shown in \autoref{eq:scale_mol_fraction}. This approach ignores excess volume of mixing, which is usually small. For example, a mixture of one mole water and one mole ethanol has an experimental density of about $28\ kmol/m^3$ \cite{Engineers-ethanol-water} which is just 3\% larger than a density estimated using the pure component molar densities in \autoref{eq:density}. 

\begin{equation} \label{eq:density}
    \rho_{solvent} = \frac{1}{\sum_j \frac{x_{solvent_j,scaled}'}{\rho_{solvent_j}}}
\end{equation}

The solvation free energy is proportional to the negative log of density. If the density estimate is off by 10\%, the resulting solvation free energy changes by about 0.06 kcal/mol ($\Delta \Delta G = 1.987 *10^{-3} \ kcal/mol K * 298 \ K * \ln{1.1} = 0.056 \ kcal/mol$).

The solvent density cannot be calculated when at least one of the solvent component's pure component liquid density cannot be calculated at the VLE measurement's temperature. This happens when either a correlation for that species' liquid density is not available in DIPPR or when the correlation that is available, but the measurement's temperature is outside the correlation's temperature range. This usually occurs when the pure species would be a solid at the measurement's temperature. 

It is assumed that the components in the solvent are miscible at the experimental composition, pressure, and temperature. These data points were labeled as VLE and not VLLE, so it is reasonable to assume there is a single liquid phase. 

    \item This was done using \autoref{eq:VLE_gsolv_final}.

    \item While comparing SolProp-mix's performance on the test sets to COSMOtherm's performance, we found a few data points where both SolProp-mix and COSMOtherm deviated from the experimental value enough to warrant a closer look at the experimental data. In the end it was decided to remove data points from two sources, \cite{suspectvle1} (three-component VLE) and \cite{suspectvle2} (four-component VLE). In addition to this, five more suspect data point was taken from the three-component VLE and one from the four-component VLE. It was also found that some data points were duplicates, usually stemming from the original VLE data being compiled in the DDB from two different sources.

    \item In some VLE measurements, one or more of the components were not observed in the liquid phase. Three-component VLE could produce both solvation free energies for both binary solvents and mono-solvents. Four-component VLE could produce solvation free energies for ternary solvents, binary solvents, and mono-solvents.

\end{enumerate}

The steps detailed in this section are visually depicted in \autoref{fig:vle_curation}. Starting with 99987 and 8835 data entries for ternary and quaternary VLE respectively, we produce a total of 29129 and 4242 solvation energies in binary and ternary solvents respectively. Data flows are reported in the diagram showing the relative impact of the different filtration steps we perform. The number of data points at each step is also detailed in \autoref{tab:vledatabases}.

\begin{table} 
\caption{The number of data points at each step of VLE filtering and conversion to solvation free energy.} \label{tab:vledatabases}
\begin{tabular}{|l|l|l|l|}
\hline
Step \# & Data point type & Three component & Four component \\
\hline
1  & VLE             & 99987 & 8835  \\ \hline
\multirow{5}{*}{2} & No T    & 2454  & 254   \\ 
   & No P            & 4693  & 471   \\ 
   & No y            & 30403 & 1577  \\
   & No x            & 351   & 10    \\ \cline{2-4}
   & VLE             & 62086 & 6523  \\ \hline
\multirow{4}{*}{3} & Metals  & 249   & 0     \\
   & Ionic liquids   & 12005 & 2132  \\
   & Non-metal salts & 49    & 0     \\ \cline{2-4}
   & VLE             & 49783 & 4391  \\ \hline
4  & VLE             & 45958 & 2922  \\ \hline
5  & VLE             & 25686 & 2035  \\ \hline
6  & VLE             & 25686 & 2035  \\ \hline
7  & Solute-Solvent  & 29855 & 4439  \\ \hline
\multirow{3}{*}{8} & No correlation  & 330   & 28    \\
   & Not in T-range  & 137   & 27    \\
   \cline{2-4}
   & Solute-Solvent  & 29396 & 4384  \\ \hline
9  & Solute-Solvent  & 29396 & 4384  \\ \hline
10 & Solute-Solvent  & 29157 & 4243  \\ \hline
\multirow{4}{*}{11} & Solute-Solvent  & 29157 & 4243  \\ \cline{2-4}
   & Mono-solvent    & 29    & 0     \\
   & Binary solvent  & 29128 & 1     \\
   & Ternary solvent & 0     & 4242  \\ \hline
\end{tabular}
\end{table}

\subsubsection{The IDAC curation process}
\begin{figure}[]
    \centering
    \includegraphics[width=0.95\textwidth]{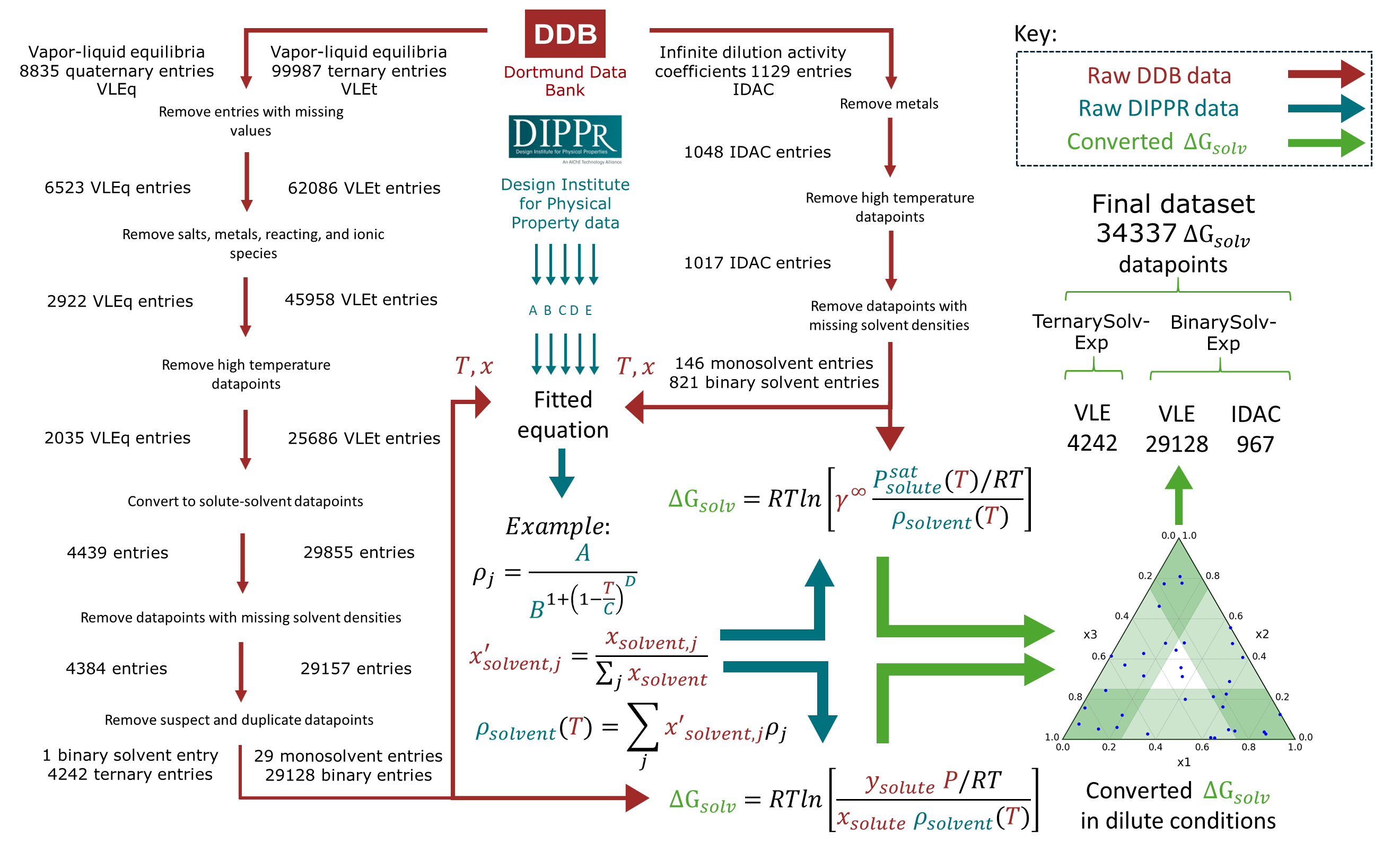}
    \caption{Detailed diagram of vapor-liquid equilibrium and infinite dilution activity coefficient data curation and conversion to solvation energies. Data filtering steps that did not reduce the number of data points are not shown.}
    \label{fig:vle_curation}
\end{figure}

The process of converting experimental infinite dilution activity coefficients in binary solvent to $\Delta G^\infty_{solv}$ was similar to that of VLE. A visual depiction of the steps in this section are shown in \autoref{fig:vle_curation}. The initial 1129 activity coefficients are filtered to produce a total of 967 solvation energies in binary solvents. The number of data points at each step is also detailed in \autoref{tab:activitydatabases}.

\begin{enumerate}
    \item Extract data from the DDB.
    \item Remove data points that include metals, ionic liquids, and non-metal salts.
    \item Remove data points where components may react with each other.
    \item Remove data points measured at temperatures above a max temperature cutoff. 
    \item Convert component DDB names to InChIs and remove any data points that fail this conversion.
    \item Calculate the solvent density for each data point and remove any data points where it cannot be calculated.
    \item Calculate the solute vapor pressure for each data point and remove any data points where it cannot be calculated.
    \item Calculate $\Delta G^\infty_{i,\,solv}$.
\end{enumerate}

The data from IDACs includes several systems where the activity coefficients were measured at the whole range of solvent compositions, from pure solvent 1 to pure solvent 2. The solvation energies at the pure solvent end points are also included in BinarySolv-Exp despite, being monosolvent data points, so that SolProp-mix's performance could be evaluated across the whole range of solvent compositions for these systems. 

\begin{table} 
\caption{The number of data points at each step of IDAC filtering and conversion to solvation free energy.} \label{tab:activitydatabases}
\begin{tabular}{|l|l|l|}
\hline
Step \# & Data point type & Binary solvent \\ \hline
1  & Activity        & 1129 \\ \hline
\multirow{4}{*}{2}   & Metals          & 81 \\
   & Ionic liquids   & 0 \\
   & Non-metal salts & 0 \\ \cline{2-3}
   & Activity        & 1048 \\ \hline
3  & Activity        & 1048 \\ \hline
4  & Activity        & 1017 \\ \hline
5  & Activity        & 1017 \\ \hline
\multirow{3}{*}{6}   & No correlation  & 0 \\
   & Not in T-range  & 50 \\
   \cline{2-3}
   & Activity        & 967 \\ \hline
7  & Activity        & 967 \\ \hline
8  & Activity        & 967 \\ \hline
\end{tabular}
\end{table}

\subsection{Experimental solvation free energy in mixed solvent dataset characterization}

\clearpage
\section{TernarySolv-Exp distribution plot}

\begin{figure}[h!]
\centering
\begin{subfigure}{.5\textwidth}
    \centering
    \includegraphics[width=\linewidth]{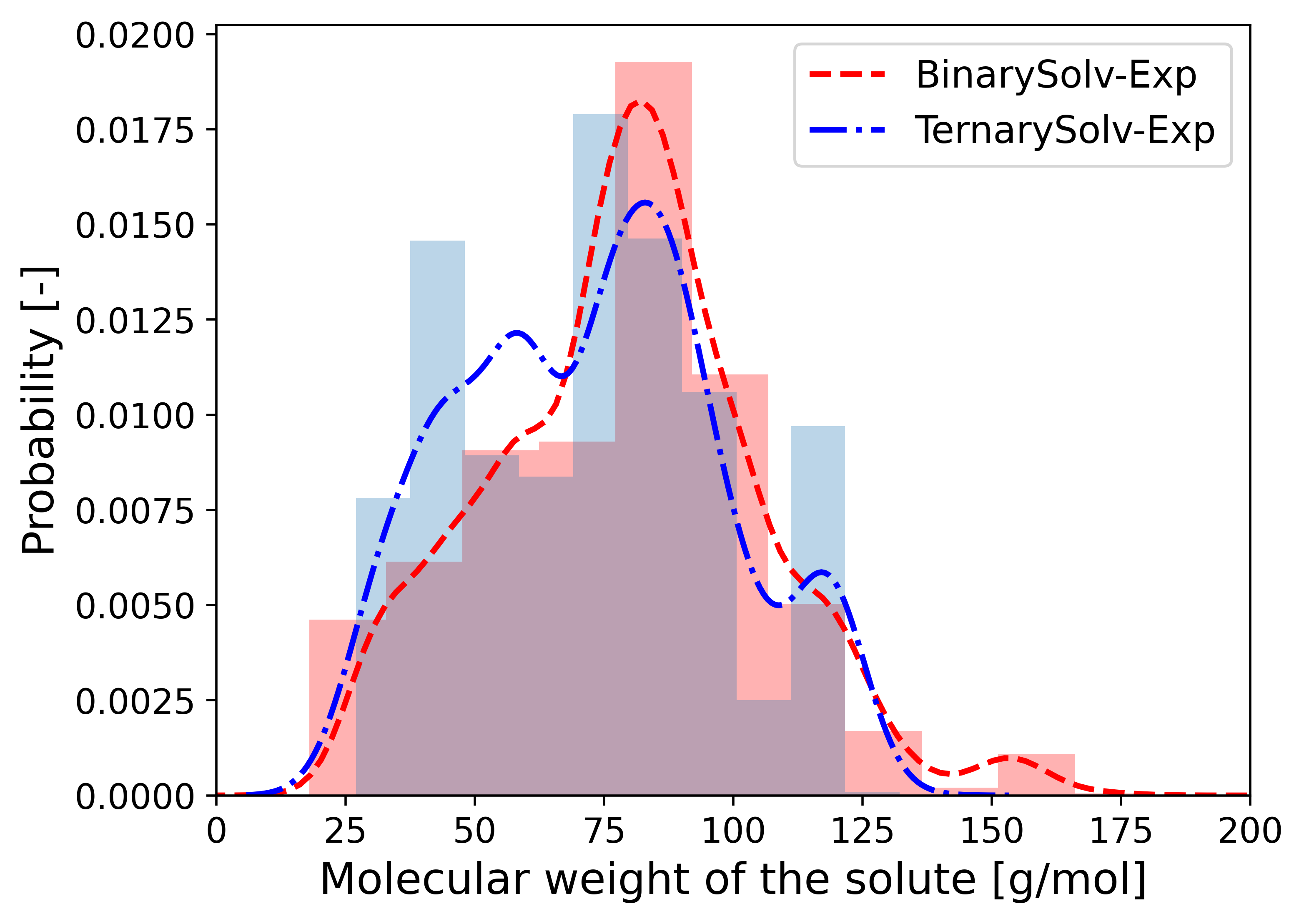}
    \caption{}
    \label{fig:1png}
\end{subfigure}%
\begin{subfigure}{.475\textwidth}
    \centering
    \includegraphics[width=\linewidth]{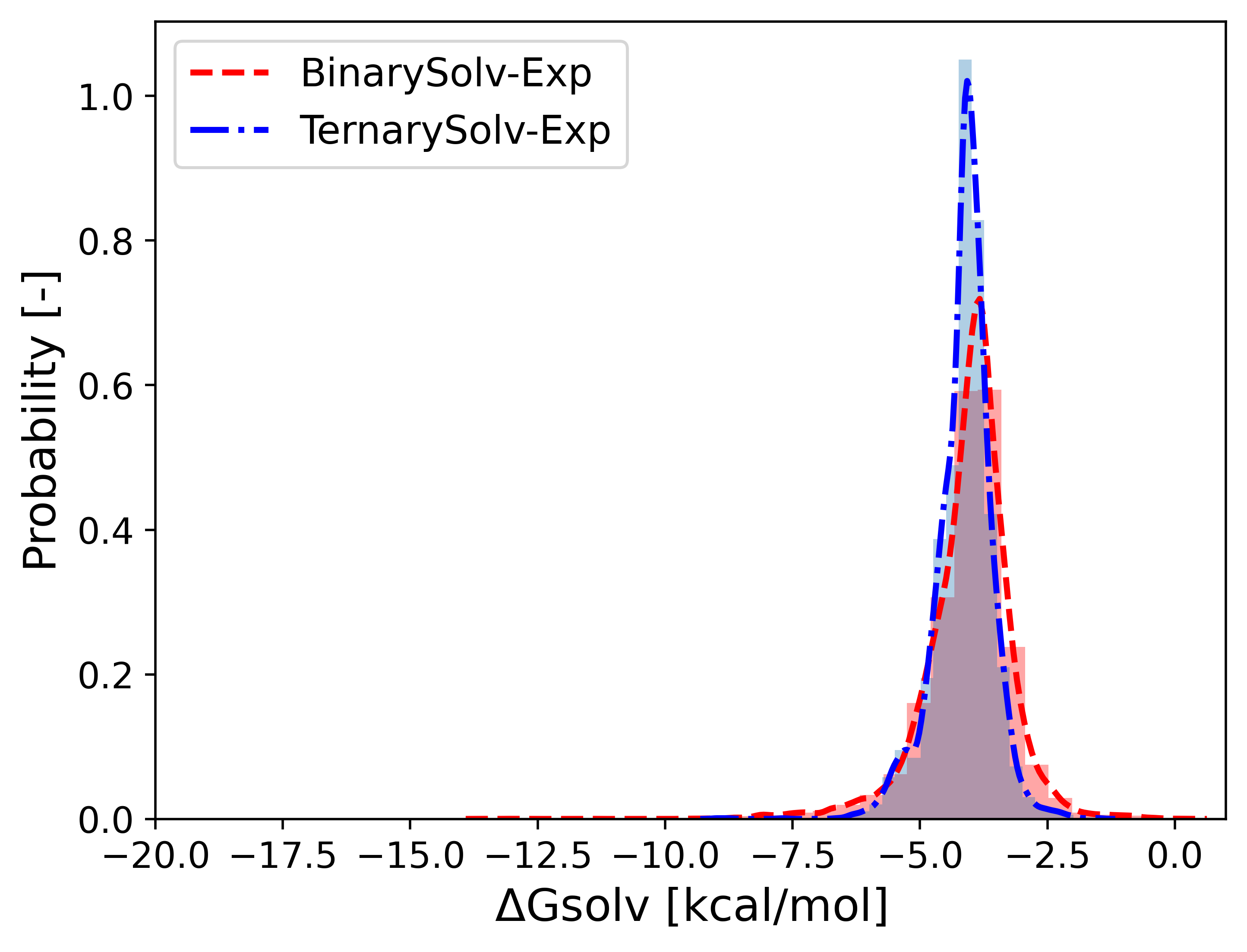}
    \caption{}
    \label{fig:3png}
\end{subfigure}
\caption{The distributions of the solute molecular weights (left) and solvation free energies (right) in the BinarySolv-Exp and TernarySolv-Exp databases.}
\label{fig:binterdist}
\end{figure}
\section{Solvation enthalpy databases}

In this part, the available databases for solvation enthalpy are discussed. CombiSolvH-Exp and CombiSolvH-QM originate from previous work of solvation free energy in monosolvent \cite{VermeireTL}. Additionally, for the same data points as CombiSolv-QM also the enthalpy of solvation was made available in CombiSolvH-QM. Comparing CombiSolvH-Exp to the QM sets, CombiSolvH-QM and BinarySolvH-QM, shows that CombiSolvH-Exp contains a high number of unique solvents, and a relatively smaller number of solutes. Subsequently, in the manuscript CombiSolv-Exp showed a shift to lighter molecules in the distributions compared to the QM sets, here a similar shift for CombiSolvH-Exp was found. Due to the similar characteristics of CombiSolv-Exp and CombiSolvH-Exp, they were merged for the purpose of the multi-task transfer learning procedure.

\begin{table}[h!]
\caption{An overview of the databases used in the training and validation of the machine learning models. The Combi abbreviation is used for monosolvents, Binary for binary solvents, Ternary for ternary solvents, Exp stands for experimental data, and QM is used for quantum chemistry databases.}
\label{tab:databases}
\small
\begin{tabular}{|l|l|l|l|l|}
\hline
\rowcolor[HTML]{EFEFEF} 
Database       & Data points & Unique solvents & Unique solutes & Temp    \\ \hline
CombiSolvH-Exp & 6,609       & 1,514           & 1,762          & $298 K$ \\ \hline
CombiSolvH-QM  & 800,000     & 284             & 10,836         & $298 K$ \\ \hline
BinarySolvH-QM & 1,000,000   & 32              & 10,960         & $298 K$ \\ \hline
\end{tabular}
\end{table}

\begin{figure}[h!]
\centering
\begin{subfigure}{.5\textwidth}
  \centering
  \includegraphics[width=\linewidth]{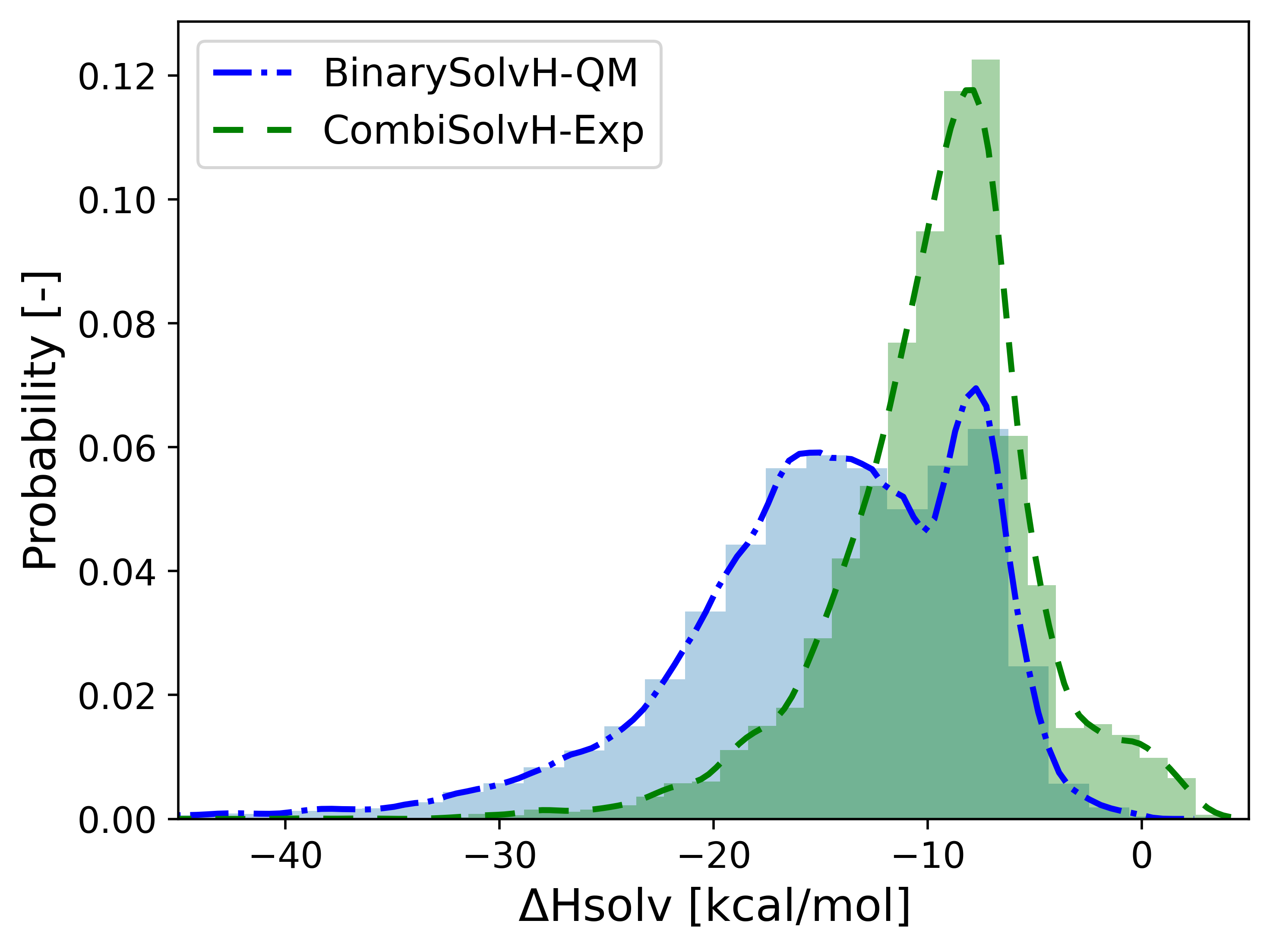}
  \caption{}
  \label{fig:sub1}
\end{subfigure}%
\begin{subfigure}{.5\textwidth}
  \centering
  \includegraphics[width=\linewidth]{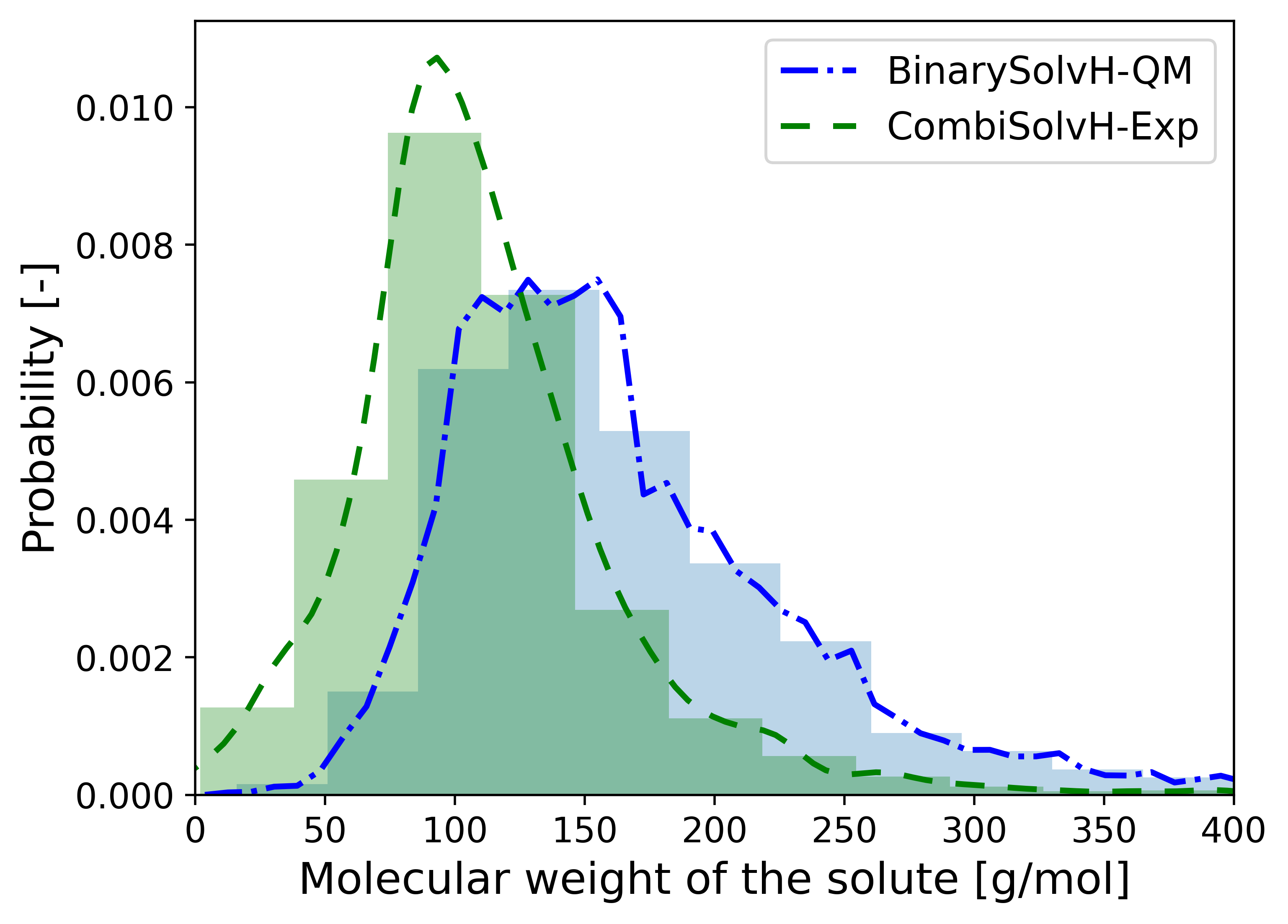}
  \caption{}
  \label{fig:sub2}
\end{subfigure}
\caption{The distributions of the solute molecular weights (a) and solvation enthalpies (b) in the BinarySolvH-QM and CombiSolvH-Exp databases.}
\label{fig:test}
\end{figure}

\newpage
\section{Hyperparameter and architecture details}

\autoref{tab:features} lists all the atom and bond features that were used for the graph representations. In a previous work, they were adapted from the standard version of Chemprop \cite{Chemprop} to make them specific to solvation related properties \cite{VermeireTL}. The values of the atom and bond features were assigned by RDKit. Additionally, molecular features, specifically the RDKit-calculated topological polar surface area and the RDKit-calculated molecular size, were concatenated with the molecular latent representations of the solvents and solute.

\begin{table}[!h]
\caption{Feature vectors for the atoms and bonds.}
\label{tab:features}
\begin{tabular}{|l|l|}
\hline
\rowcolor[HTML]{EFEFEF} 
\textbf{Atom feature vectors}               & \textbf{Bond feature vectors} \\ \hline
Atomic number                      & Bond type            \\ \hline
Number of neighbouring atoms       & Conjugation          \\ \hline
Formal charge                      & Ring type            \\ \hline
Number of connected hydrogen atoms & Stereo-chemistry     \\ \hline
Hybridization                      &                      \\ \hline
Number of lone pairs               &                      \\ \hline
Hydrogen bond donating character   &                      \\ \hline
Hydrogen bond accepting character  &                      \\ \hline
Ring size                          &                      \\ \hline
Aromaticity                        &                      \\ \hline
Electronegativity                  &                      \\ \hline
Atomic molar mass                  &                      \\ \hline
\end{tabular}
\end{table}

Since this manuscript focused on the application of a pooling function for mixtures, rather than the optimization of the neural networks, many of the hyperparameters for the model architecture and optimization of the neural network were fixed. The hyperparameters were kept constant with previous work, where hyperparameters optimization studies were performed \cite{VermeireTL}. For the D-MPNN, the depth of the message passing was set to 4 and the size of the hidden layers to 200 for both the solvents and solute embeddings. The D-MPNN linear layers had no bias and the results were obtained without considering dropout. A LeakyReLU activation function was used to connect the different layers of the neural network. For the FFN, 4 layers were considered each with a hidden size of 500. The linear layers had a bias, no dropout, and were also connected with a LeakyReLU activation function. Prior to training the neural network, both targets were normalized using the standard score. The parameters of the neural network were initialized randomly by a normal distribution as published by Glorot et al. \cite{Glorot_Bengio_2010}, except in the case of transfer learning where the model parameters were initialized using parameters from the pre-trained neural network. Training of the neural network was performed in batches of 50 data points for 10 epochs. A Noam learning rate scheduler was used with piece-wise linear increase and exponential decay, based on the learning rate scheduler in the Transformer model for Natural Language Processing \cite{attentionneed}. The model parameters were optimized with stochastic optimization as implemented in the Adam algorithm \cite{adam} and based on the mean-squared-error loss. 

\section{Additional trends plots for binary solvent systems}

\begin{figure}[h!]
\centering
\begin{subfigure}{.5\textwidth}
  \centering
  \includegraphics[width=\linewidth]{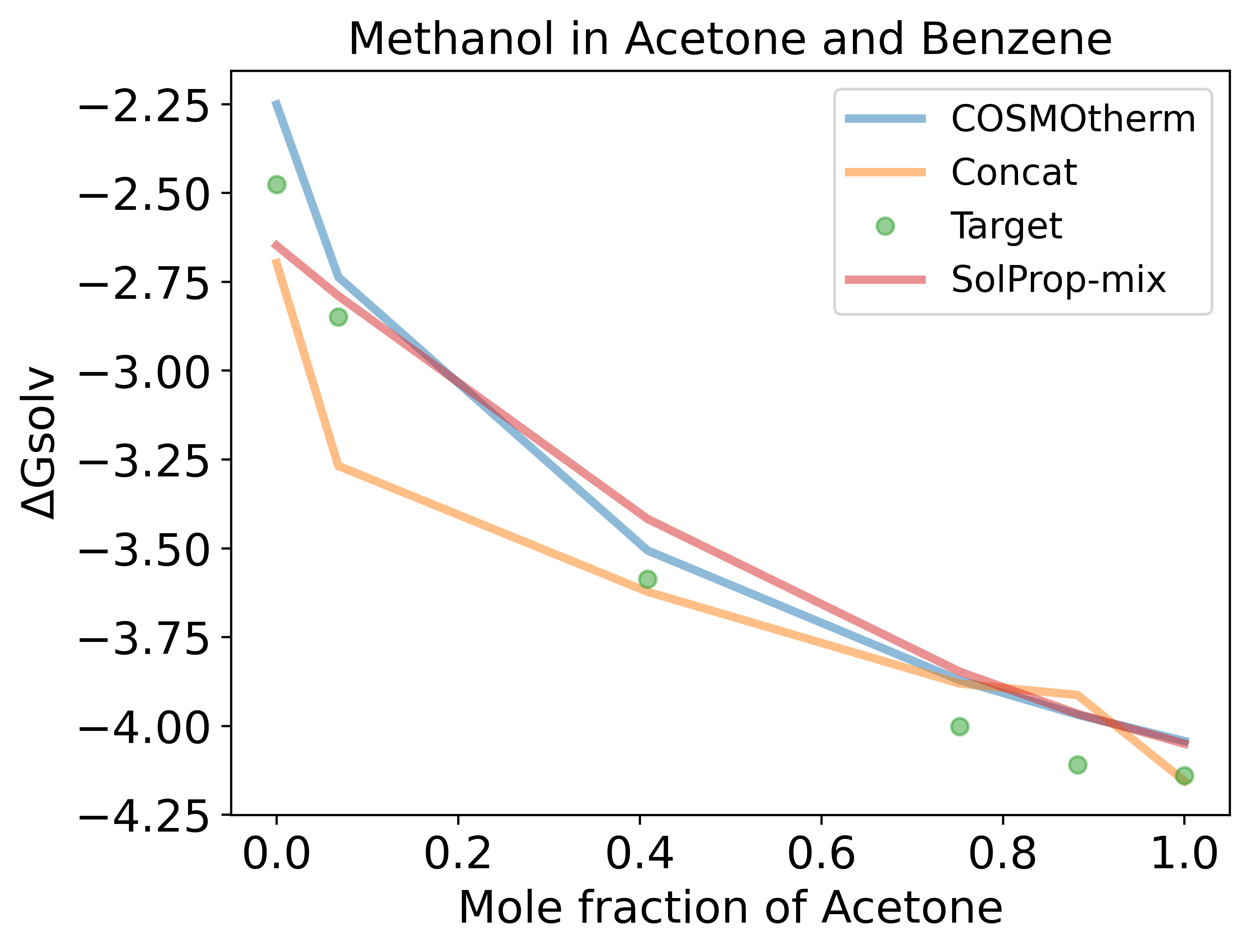}
  \caption{}
  \label{}
\end{subfigure}%
\begin{subfigure}{.5\textwidth}
  \centering
  \includegraphics[width=\linewidth]{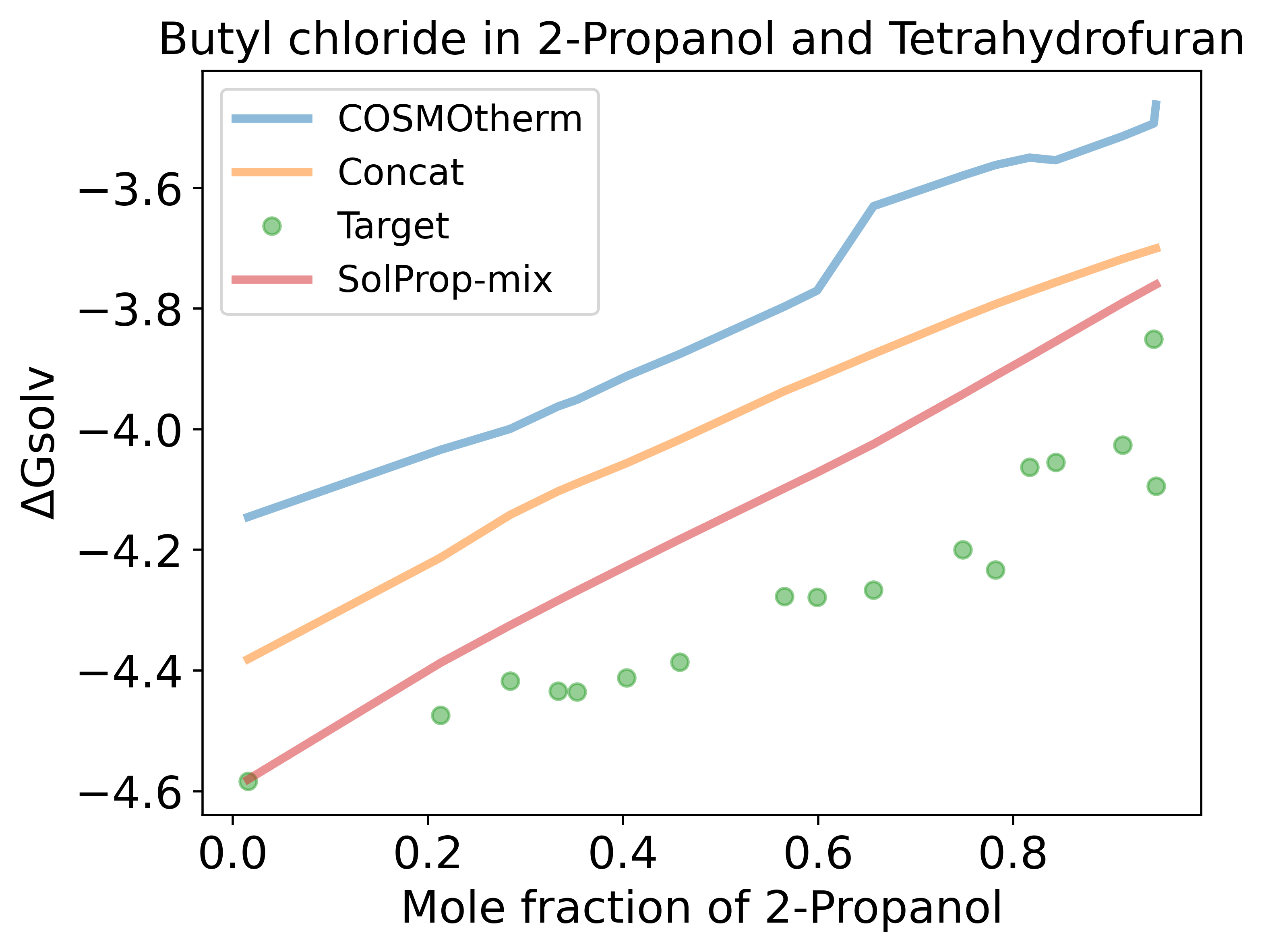}
  \caption{}
  \label{}
\end{subfigure}
\caption{}
\label{fig:test}
\end{figure}

\begin{figure}[h!]
\centering
\begin{subfigure}{.5\textwidth}
  \centering
  \includegraphics[width=\linewidth]{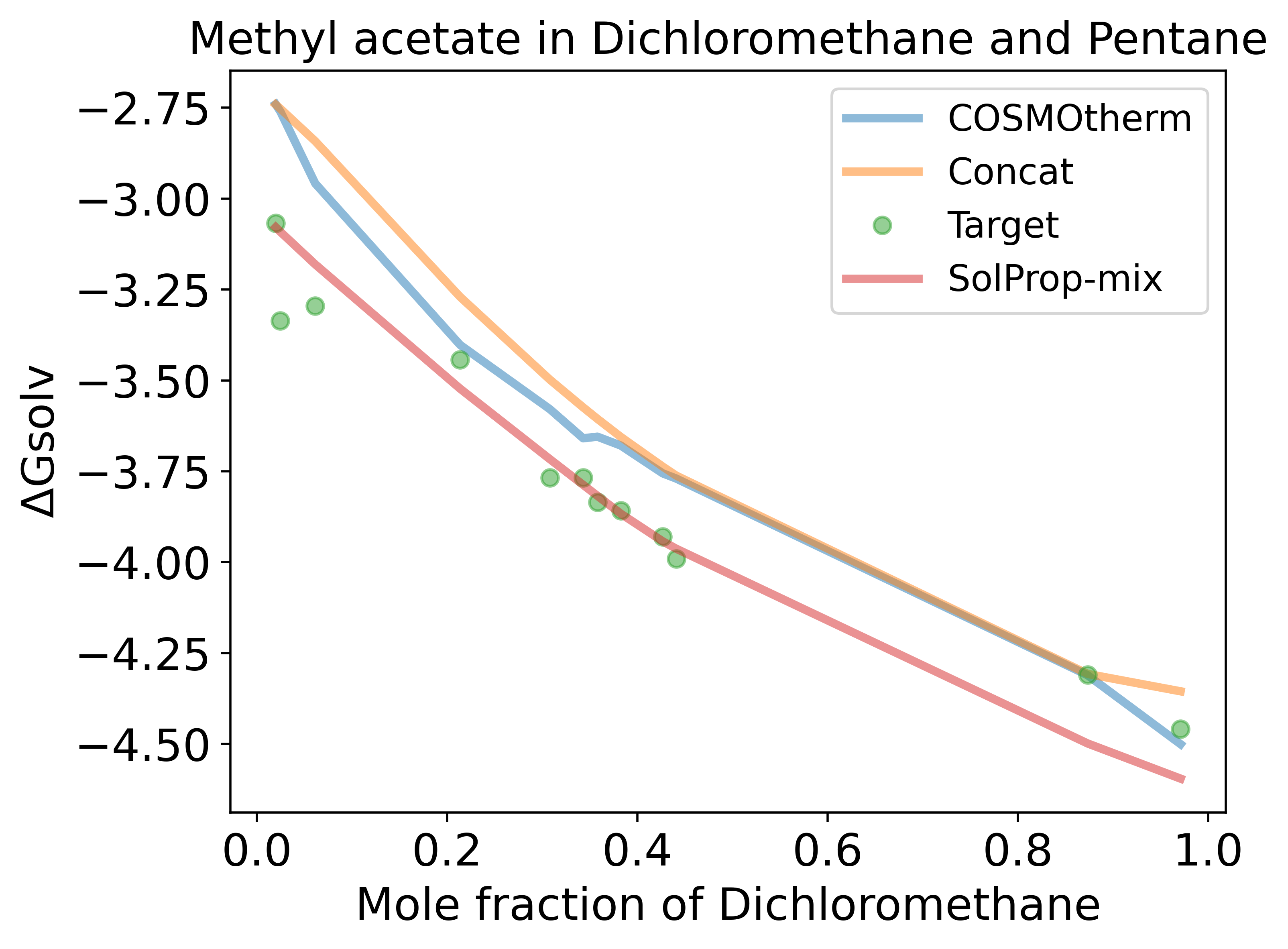}
  \caption{}
  \label{}
\end{subfigure}%
\begin{subfigure}{.5\textwidth}
  \centering
  \includegraphics[width=\linewidth]{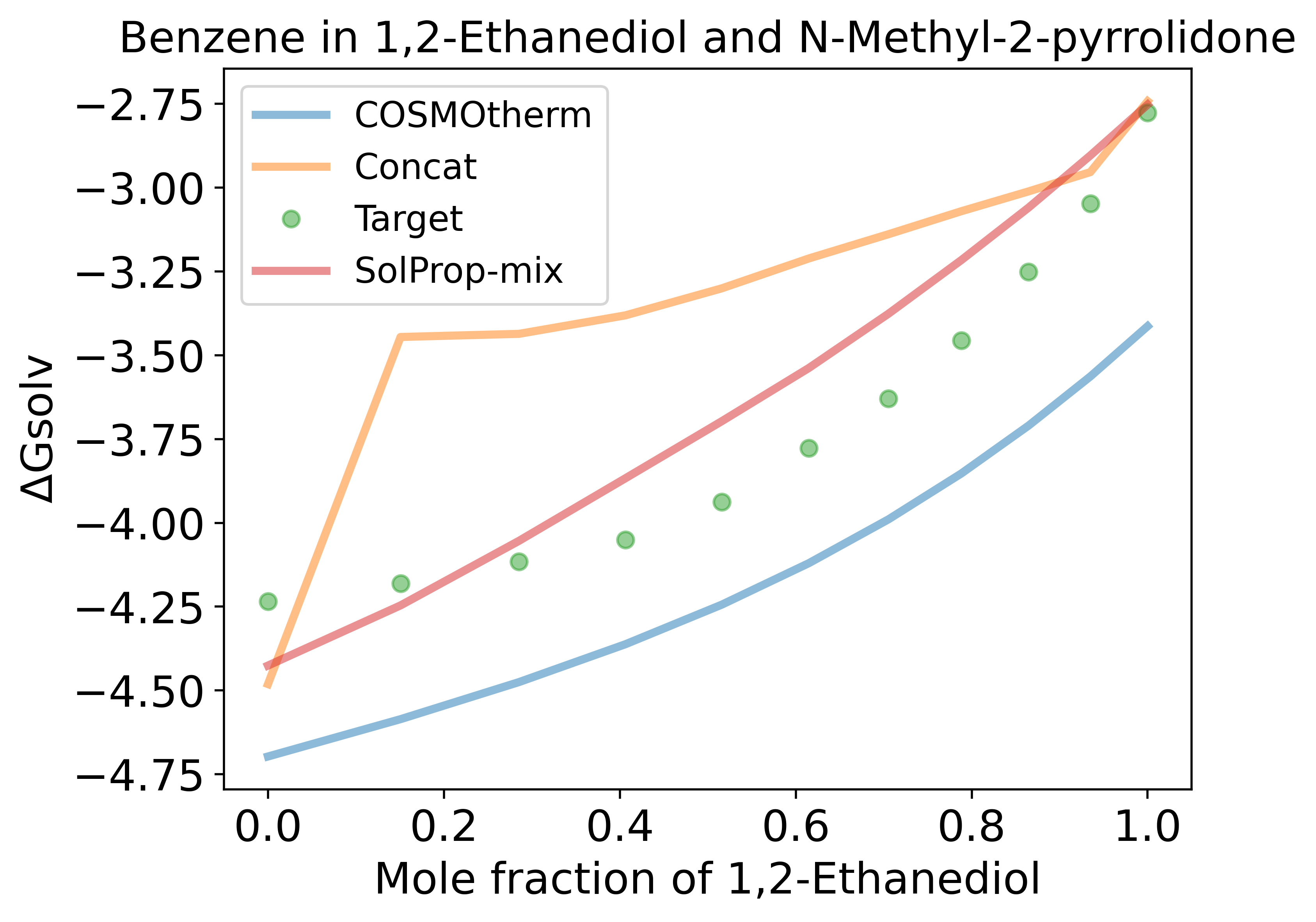}
  \caption{}
  \label{}
\end{subfigure}
\caption{}
\label{fig:test}
\end{figure}

\begin{figure}[h!]
\centering
\begin{subfigure}{.5\textwidth}
  \centering
  \includegraphics[width=\linewidth]{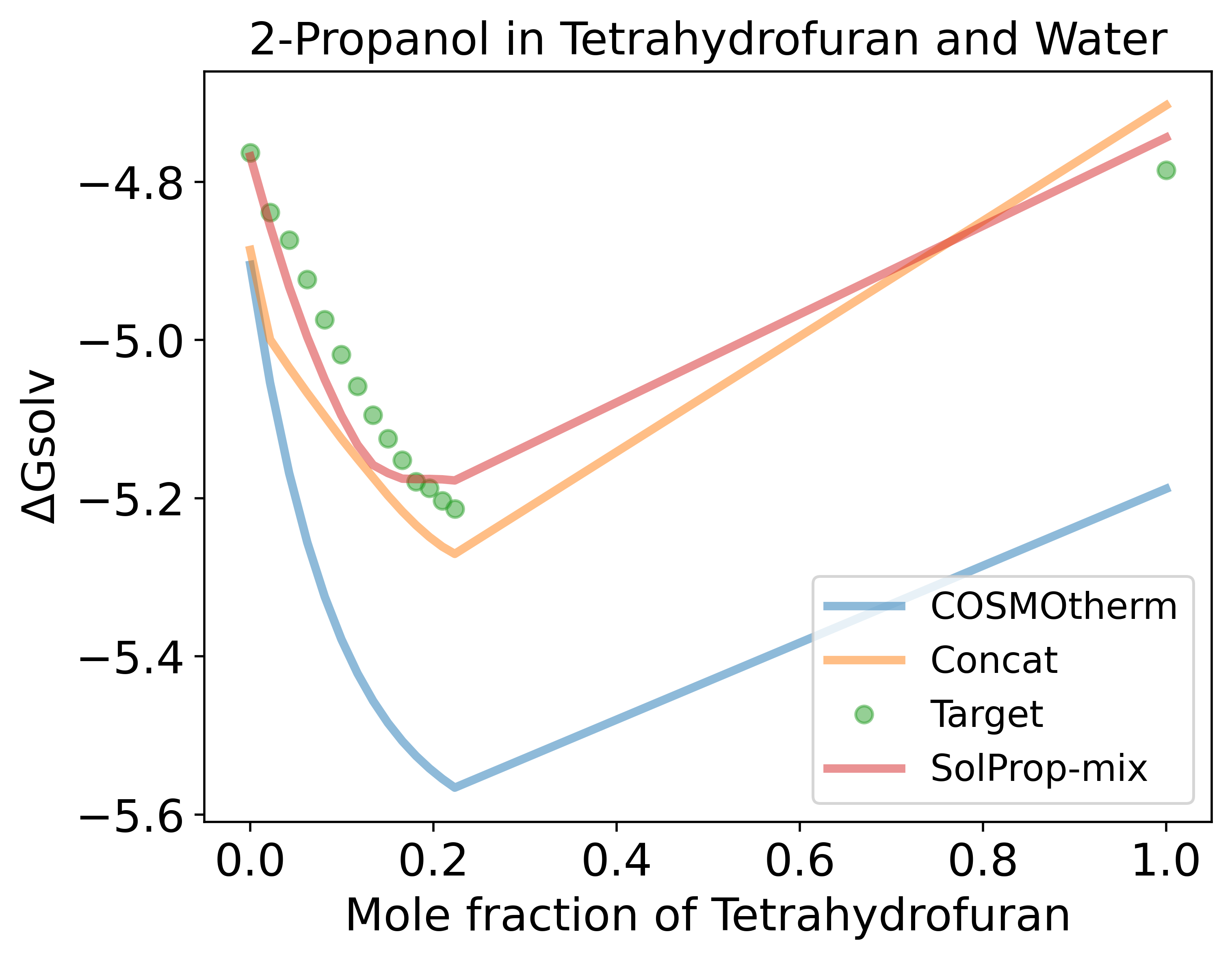}
  \caption{}
  \label{}
\end{subfigure}%
\begin{subfigure}{.5\textwidth}
  \centering
  \includegraphics[width=\linewidth]{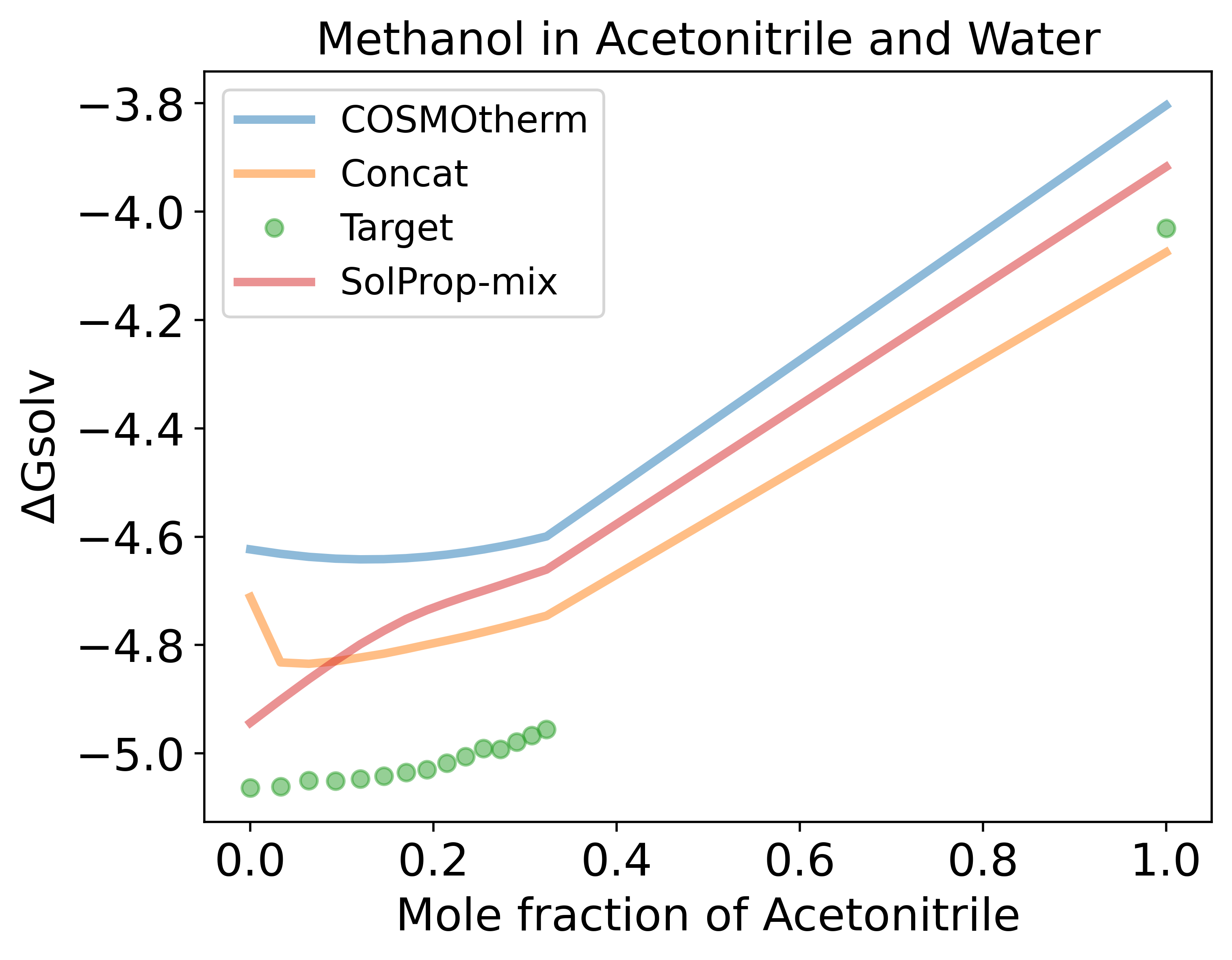}
  \caption{}
  \label{}
\end{subfigure}
\caption{}
\label{fig:test}
\end{figure}

\begin{figure}[h!]
\centering
\begin{subfigure}{.5\textwidth}
  \centering
  \includegraphics[width=\linewidth]{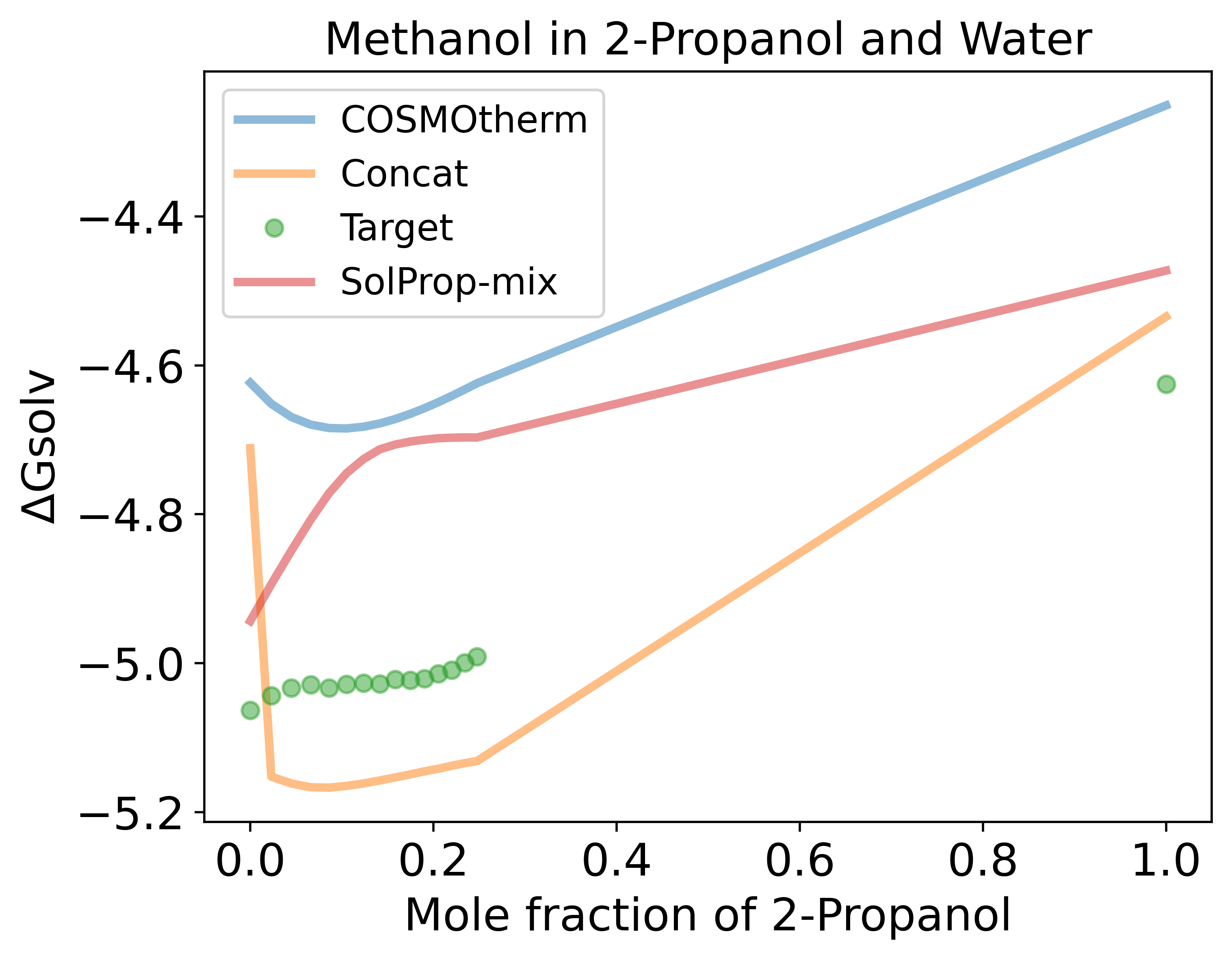}
  \caption{}
  \label{}
\end{subfigure}%
\begin{subfigure}{.5\textwidth}
  \centering
  \includegraphics[width=\linewidth]{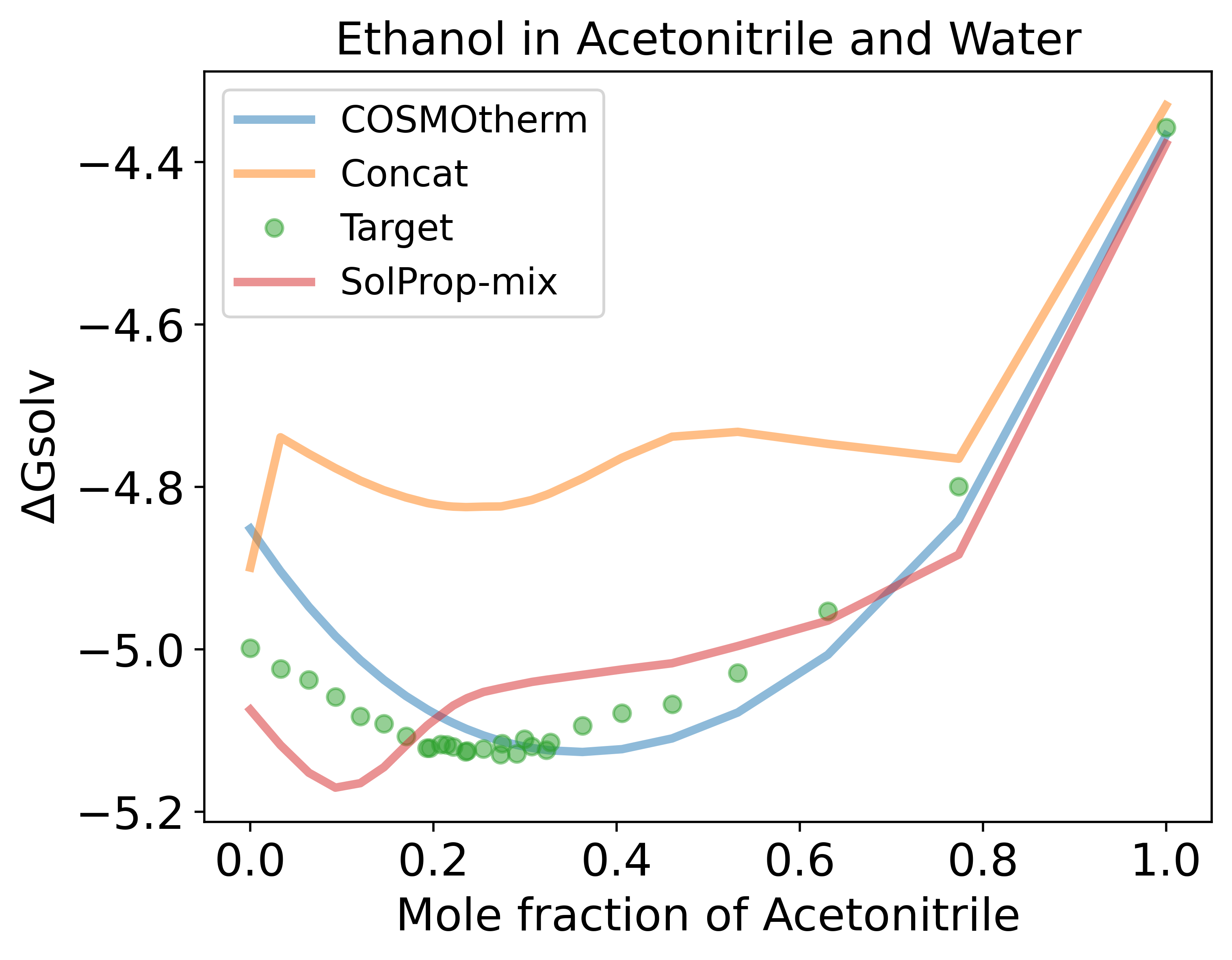}
  \caption{}
  \label{}
\end{subfigure}
\caption{}
\label{fig:test}
\end{figure}
\clearpage
\clearpage

\bibliographystyle{elsarticle-num} 
\bibliography{achemso-supl}